\documentclass{article}

\usepackage[main,final]{neurips_2026}

\usepackage[utf8]{inputenc}
\usepackage[T1]{fontenc}
\usepackage{float}
\usepackage{hyperref}
\usepackage{url}
\usepackage{booktabs}
\usepackage{amsfonts}
\usepackage{nicefrac}
\usepackage{microtype}
\usepackage{amsmath}
\usepackage{graphicx}
\usepackage{doi}
\usepackage{placeins}
\usepackage{booktabs}
\usepackage{multirow}
\usepackage{tabularx}
\usepackage{makecell}
\usepackage{array}
\usepackage[table]{xcolor}


\hypersetup{
    colorlinks=true,
    linkcolor=blue,
    citecolor=blue,
    urlcolor=blue
}

\title{\bfseries
AirMPA: A Meteorology-to-Pollution Adapter for Global Air Quality Forecasting
}

\author{
\textbf{Yiheng Wang}$^{1,2}$,
\textbf{Kai Zheng}$^{2}$,
\textbf{Yuetan Lin}$^{2}$,
\textbf{Fanglu Fan}$^{1,2}$, \\
\textbf{Chenliang Tao}$^{1}$,
\textbf{Guochao Chen}$^{1}$,
\textbf{Hongliang Zhang}$^{3,*}$,
\textbf{Hao Li}$^{1,2,*}$ \\[6pt]
{\small $^{1}$Fudan University
\quad $^{2}$Shanghai Academy of AI for Science}\\
{\small $^{3}$University of Shanghai for Science and Technology}
}

\begin{document}

\maketitle
\begin{abstract}

Accurate real-time forecasting of atmospheric pollutants is essential for reducing exposure risks and improving public health. Traditional numerical forecasting systems suffer from incomplete parameterizations, uncertain inputs, and high computational cost. Existing deep learning approaches improve efficiency, but often entangle meteorological and pollutant variables within unified forecasting frameworks, limiting flexibility and pollutant-specific representation learning. To address these challenges, we propose AirMPA, a decoupled autoregressive framework for meteorology-to-pollution forecasting. AirMPA predicts 13 atmospheric pollutants at 0.4$^\circ$ spatial resolution and 12-h temporal resolution using meteorological fields, emission priors, and historical pollutant concentrations. The model is built on an improved Swin-Transformer architecture with zonal release attention and dual-branch fusion, enabling more effective modeling of global transport continuity and heterogeneous pollutant dynamics. During training, AirMPA uses the fifth-generation atmospheric reanalysis from the European Centre for Medium-Range Weather Forecasts (ERA5) as meteorological forcing, while during inference it directly ingests forecast fields from external meteorological models, enabling stable autoregressive prediction up to 5 days ahead. Experiments show that AirMPA consistently outperforms Aurora and surpasses the Copernicus Atmosphere Monitoring Service (CAMS) in approximately 90\% of pollutant--lead-time evaluation cases over the 5-day forecast horizon, while outperforming CAMS on all variables beyond 48 h. These results demonstrate the potential of decoupled meteorology-to-pollution forecasting for global air quality prediction.
\end{abstract}

\section{Introduction}
Air pollution has become a major global public health threat, causing millions of premature deaths each year and imposing substantial socioeconomic burdens \cite{tao2025inequality,zhan2023impacts,lelieveld2015contribution}. Accurate multi-day forecasting of atmospheric pollutants is therefore essential for reducing exposure risks, supporting environmental management, and mitigating extreme pollution events\cite{masood2021review,Subramaniam2022AI}.

Air quality forecasting is intrinsically challenging because atmospheric pollutant evolution is governed by highly nonlinear interactions among meteorology, chemistry, and emissions, making it difficult to simultaneously achieve high accuracy, computational efficiency, and long-term stability \cite{liu2020airpollution, cha2025investigating}. Traditional numerical models, such as WRF-Chem and CMAQ ~\cite{zhang2025uiwrfchem,pan2026mpascmaq}, can explicitly simulate complex physical and chemical processes, but their practical use in real-time forecasting is limited by high computational cost as well as uncertainties in emission inventories and chemical parameterizations \cite{ojha2020indogangetic, zhang2025major,knowland2022geoscf}. Data-driven approaches have substantially improved computational efficiency and shown strong potential for atmospheric forecasting~\cite{ding2026diffusion,kheder2026topoflow,liang2023airformer,bodele2025advanced}. Recent studies have further demonstrated that air pollution prediction can be incorporated into unified meteorological forecasting frameworks, achieving performance competitive with traditional systems such as the Copernicus Atmosphere Monitoring Service (CAMS)~\cite{gui2026advancing,bodnar2025foundation}. However, these methods usually couple meteorological and pollutant variables at the data level, which reduces modeling flexibility and may weaken the representation of pollutant-specific dynamics, thereby limiting forecasting accuracy and transferability across meteorological systems.

Meanwhile, the rapid advancement of meteorological foundation models provides new opportunities for addressing this challenge. High-accuracy forecasting systems include the Integrated Forecasting System (IFS) of the European Centre for Medium-Range Weather Forecasts (ECMWF) and data-driven models such as the Artificial Intelligence Forecasting System (AIFS), FuXi, Pangu-Weather, and GraphCast. These systems have achieved substantial progress in medium- to long-range weather forecasting, providing high-resolution and high-fidelity atmospheric fields.~\cite{lang2024aifs,chen2023FuXi,zhong2024FuXi2,bi2023accurate,lam2023learning}. In traditional numerical models, pollutant prediction is inherently driven by meteorological processes \cite{wong2024development,sun2026development}. This paradigm suggests the possibility of directly using meteorological forecasts to drive downstream air pollutant prediction.

In this work, we propose AirMPA, a decoupled global air pollutant forecasting framework that explicitly separates meteorological forecasting from pollutant prediction. AirMPA produces 5-day forecasts of 13 atmospheric pollutants at $0.4^\circ$ spatial resolution and 12\,h temporal resolution. Our main contributions are as follows:
\begin{itemize}
    \item We propose AirMPA, a decoupled meteorology-to-pollution forecasting framework that treats pollutant prediction as an independent downstream task driven by external meteorological forcing, producing global 12-hourly forecasts at $0.4^\circ$ resolution.

    \item The decoupled design enables a clean transfer from reanalysis-driven training to forecast-driven inference, providing a practical pathway for using external meteorological forecasts in downstream air pollutant prediction.
    
    \item The improved Swin-Transformer architecture, equipped with zonal release attention and dual-branch historical fusion, better captures the continuity of global atmospheric pollutant transport and the heterogeneity of pollutant temporal evolution, thereby improving forecasting accuracy.

    \item Validation against station observations and case studies of global ozone pollution events further demonstrate the practical significance of AirMPA.
\end{itemize}

\section{Related Work}

\subsection{Numerical Models}
CAMS, operated by ECMWF, provides global atmospheric forecasts twice daily \cite{inness2019cams}. By integrating satellite observations and ground-based measurements through data assimilation, CAMS yields consistent initial states to drive physical and chemical transport models. The system outputs over 50 chemical species and 7 aerosol types (e.g., dust, black carbon, and sulfates) at a standard horizontal resolution of approximately $0.4^\circ \times 0.4^\circ$, covering a 5-day forecast horizon. These comprehensive datasets serve as a foundational benchmark for global environmental governance, supporting both real-time air quality management and long-term climate change research.

\subsection{Data-Driven Deep Learning Models}
Recent advances in deep learning have improved air pollution forecasting across station, regional, and global scales \cite{liang2023airformer,geng2025FuXi,tian2024air,abeysinghe2025syncast,khatibi2025advancing,harder2026aifs,nguyen2023climax}. At the station scale, AirFormer combines deterministic spatiotemporal attention with stochastic latent variables for nationwide forecasting \cite{liang2023airformer}. For regional gridded prediction, TopoFlow incorporates wind-guided patch reordering and topography-aware attention to represent transport and terrain effects \cite{kheder2026topoflow}. Air-DualODE integrates open-system physical dynamics with a complementary data-driven branch, improving prediction across city and national scales \cite{liang2025air}.

At the global scale, Aurora employs a flexible encoder-decoder architecture with a 3D Swin-Transformer backbone, delivering forecasts at $0.4^\circ$ resolution that outperform conventional chemical transport models in 74\% of evaluation metrics \cite{bodnar2025foundation}. AI-GAMFS provides multi-day global aerosol forecasts at 3-h resolution in under one minute, reducing dust concentration errors by 34\%--74\% compared to regional physical models \cite{gui2026advancing}. More recently, AIFS-COMPO jointly predicts meteorological and composition variables at approximately 80-km and 3-h resolution. It matches or improves upon CAMS for several species against independent observations, but remains less accurate for total-column ozone (O$_3$) and near-surface nitrogen dioxide (NO$_2$) against CAMS analysis \cite{harder2026aifs}.

While coupled global systems jointly forecast meteorology and pollutants, Fan et al.~\cite{fan2025offline} explore offline meteorology--pollution coupling through bilinear pooling. AirMPA also adopts a decoupled formulation, with explicit emission conditioning, zonal release attention, and pollutant-specific historical fusion. It treats pollutant forecasting as a dedicated downstream task and can incorporate forecasts from different meteorological systems without retraining the pollutant model.

\section{Meteorology to Pollutant Adapter}

\subsection{Problem Formulation}
To improve modularity and enable direct reuse of high-quality meteorological forecasts, we formulate global air pollutant forecasting as a meteorology-driven autoregressive prediction problem. Instead of jointly modeling meteorology and atmospheric composition within a unified framework, AirMPA decouples meteorological forecasting from pollutant prediction and treats the latter as a downstream task conditioned on external meteorological fields, emission priors, and historical pollutant states.

During training, we assume access to meteorological sequences $\mathbf{M}_{0:T}$, emission sequences $\mathbf{E}_{0:T}$, and pollutant concentration sequences $\mathbf{P}_{0:T}$. Here, $\mathbf{M}_t \in \mathbb{R}^{C_m \times H \times W}$, $\mathbf{E}_t \in \mathbb{R}^{C_e \times H \times W}$, and $\mathbf{P}_t \in \mathbb{R}^{C_p \times H \times W}$ denote the meteorological fields, emission priors, and pollutant concentrations at time step $t$, respectively, where $C_m$, $C_e$, and $C_p$ are the channel numbers and $H$, $W$ are the spatial dimensions on the $0.4^\circ$ global grid. In this study, the meteorological inputs are taken from ERA5 reanalysis during training and replaced by forecast fields from FuXi-2.0 during inference. Under this formulation, the meteorological fields at the target prediction time serve as forcing inputs for pollutant forecasting, driving transport, diffusion, and chemical evolution. To reduce error accumulation and preserve short-term pollutant memory, the prediction at time step $t$ is conditioned on the two preceding pollutant states, the meteorological and emission inputs at time steps $t-1$ and $t$, and the temporal encoding of the target prediction time ($\boldsymbol{\tau}_{t}$):
\begin{equation}
\hat{\mathbf{P}}_{t}
=
f_{\theta}\!\left(
\mathbf{P}_{t-2},\, \mathbf{P}_{t-1},\,
\mathbf{M}_{t-1},\, \mathbf{M}_{t},\,
\mathbf{E}_{t-1},\, \mathbf{E}_{t},\,
\boldsymbol{\tau}_{t}
\right),
\end{equation}

Under this formulation, AirMPA serves as a plug-and-play pollution forecasting module that enables a clean transition from reanalysis-driven training to forecast-driven inference.

\subsection{AirMPA Overview}

\begin{figure}[htbp]
  \centering
  \includegraphics[width=1\textwidth]{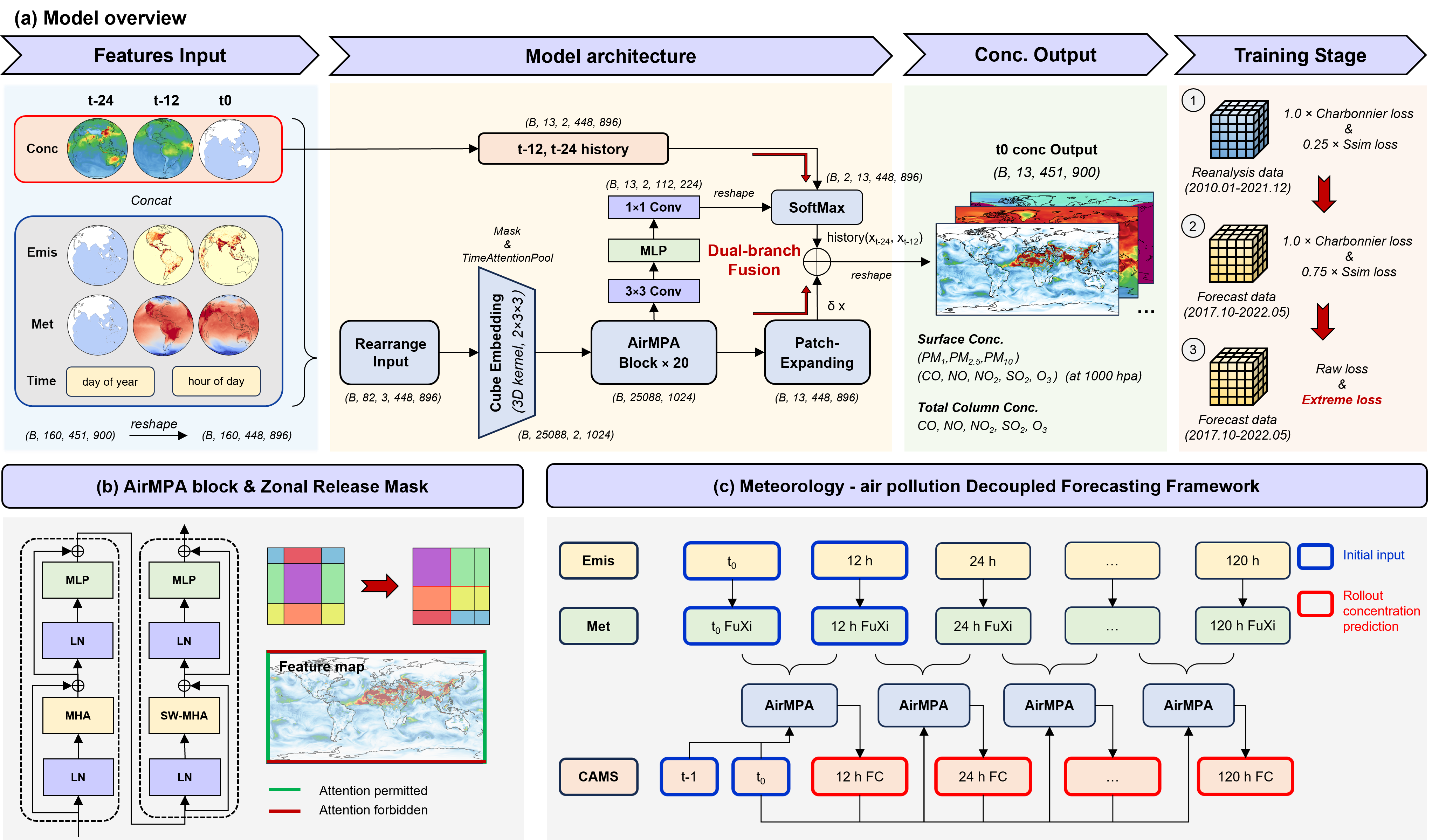}
  \caption{AirMPA architecture. (a) Overview of the AirMPA architecture and data flow. (b) Illustration of the zonal release attention mechanism, which mitigates the artificial discontinuity at the global longitude boundary. (c) Autoregressive inference pipeline for multi-step pollutant forecasting.}
  \label{fig:overview}
\end{figure}

AirMPA is built upon an improved Swin Transformer backbone for global atmospheric pollutant forecasting \cite{liu2021swin}. Given meteorological fields, emission priors, and historical pollutant concentrations, the model predicts future pollutant states in an autoregressive manner. As illustrated in Figure~\ref{fig:overview}(a), the overall framework follows the pipeline of \textit{input adaptation $\rightarrow$ cube embedding $\rightarrow$ spatiotemporal feature extraction $\rightarrow$ dual-branch pollutant fusion $\rightarrow$ multi-pollutant prediction}. To ensure compatibility with the backbone, the native global grid is first resized from $900 \times 451$ to $896 \times 448$ using bilinear interpolation, and the predicted fields are restored to the original resolution after decoding.

To encode short-term temporal dependencies before spatial feature extraction, AirMPA introduces a temporal cube embedding module. Let $\tilde{\mathbf{X}}_t^{(i)} \in \mathbb{R}^{C_i \times H \times W}$ denote the input variables at the $i$-th temporal slice. Stacking all slices and applying temporal convolution give
\begin{equation}
\tilde{\mathbf{X}}_t^{\mathrm{seq}} \in \mathbb{R}^{B \times T \times C \times H \times W},
\qquad
\mathbf{F}^{\mathrm{temp}} = \mathrm{Conv}_{3\mathrm{D}}(\tilde{\mathbf{X}}_t^{\mathrm{seq}}),
\end{equation}
where $B$, $T$, and $C$ denote the batch size, number of temporal inputs, and total number of channels, respectively.

The resulting features are then aggregated by attention pooling to obtain a compact embedded representation:
\begin{equation}
\mathbf{F}^{\mathrm{emb}} = \sum_{i=1}^{T} \alpha_i \mathbf{F}^{\mathrm{temp}}_{i}, 
\qquad
\alpha_i = \frac{\exp(s_i)}{\sum_{j=1}^{T}\exp(s_j)},
\end{equation}
where $s_i$ and $\alpha_i$ denote the learnable attention score and normalized weight of the $i$-th temporal slice. This module summarizes multi-source, multi-step inputs into a compact spatiotemporal representation for downstream forecasting.

The embedded features are then processed by stacked AirMPA blocks for spatiotemporal feature extraction. Built on this backbone, AirMPA incorporates two task-specific designs: (1) zonal release attention and (2) dual-branch fusion, and outputs forecasts for all pollutants at the current step.

\subsubsection{Zonal Release Attention}

The backbone of AirMPA is built upon shifted-window Swin Transformer blocks, which were originally developed for image processing under a planar-grid assumption with non-periodic boundaries. However, directly applying the standard shifted-window masking strategy to global latitude--longitude grids introduces an artificial discontinuity at the longitude boundary. On the Earth system grid, the left and right boundaries are physically adjacent due to the periodic topology of the globe in longitude, whereas the original masking mechanism treats them as disconnected regions. Consequently, information exchange across the dateline is unnecessarily blocked, which may hinder the representation of zonal continuity and trans-boundary transport, especially for large-scale and long-lived pollutants.

To address this mismatch, we introduce a Zonal Release Attention mechanism (Figure ~\ref{fig:overview}). The key idea is to relax the boundary constraint only along the longitude dimension while preserving the standard shifted-window masking strategy along the latitude dimension. Specifically, tokens located at the two opposite longitude edges are grouped into the same attention region and computed without masking. Let $\mathbf{M}_{ij}$ denote the original shifted-window attention mask. We define a modified mask $\mathbf{M}^{\ast}_{ij}$ as

\begin{equation}
\mathbf{M}^{\ast}_{ij} =
\begin{cases}
0, & \text{if tokens } i \text{ and } j \text{ are grouped across the longitude boundary}, \\
\mathbf{M}_{ij}, & \text{otherwise},
\end{cases}
\end{equation}
where the zero mask indicates that no artificial separation is imposed for cross-boundary token pairs.

The resulting attention is computed as
\begin{equation}
\mathrm{Attn}^{\ast}(\mathbf{Q}, \mathbf{K}, \mathbf{V})
=
\mathrm{Softmax}\left(
\frac{\mathbf{Q}\mathbf{K}^{\top}}{\sqrt{d}} + \mathbf{B} + \mathbf{M}^{\ast}
\right)\mathbf{V},
\end{equation}
where $\mathbf{Q}$, $\mathbf{K}$, and $\mathbf{V}$ denote the query, key, and value matrices, $d$ is the channel dimension, and $\mathbf{B}$ is the relative position bias. Compared with standard shifted-window attention, this modification does not increase the overall architectural complexity, but allows physically adjacent regions near the longitude boundaries to interact directly within the same attention window. In this way, the model better respects the periodic geometry of the Earth system and can more faithfully capture cross-boundary transport patterns in global atmospheric composition fields.

\subsubsection{Dual-branch Fusion for Multi-pollutant Forecasting}

Although atmospheric pollutant evolution is continuous in time, different pollutants exhibit markedly different temporal response characteristics under the 12-hour forecasting interval used in this work \cite{zhao2016annual,yang2025two,geng2025FuXi}. For example, O$_3$ and nitric oxide (NO) are strongly influenced by photochemical cycles and solar radiation, whereas fine particulate matter (PM$_{2.5}$) and sulfur dioxide (SO$_2$) are often more strongly constrained by persistence, transport accumulation, and emission carryover. Therefore, directly using a shared temporal fusion strategy for all pollutants may be suboptimal, because it implicitly assumes homogeneous temporal inheritance patterns across variables. To address this issue, we design a dual-branch fusion module that combines two complementary information sources: the latent spatiotemporal features extracted by the Transformer backbone and the historical pollutant reference fields that provide direct temporal guidance. Let

\begin{equation}
\mathbf{F}^{\mathrm{back}} \in \mathbb{R}^{B \times C_p \times H \times W}, \qquad
\mathbf{X}^{\mathrm{ref}} \in \mathbb{R}^{B \times C_p \times T_r \times H \times W},
\end{equation}

denote the backbone output for pollutant prediction and the historical pollutant reference tensor constructed from previous concentration fields, respectively, where $C_p$ is the number of target pollutants and $T_r$ is the number of reference steps. In this study, $T_r = 2$, corresponding to $\mathbf{P}_{t-2}$ and $\mathbf{P}_{t-1}$.

Instead of assigning fixed temporal weights, the historical-reference branch learns spatially adaptive and pollutant-specific weighting coefficients for different reference states. The fused reference feature is defined as
\begin{equation}
R_{b,c,h,w}
=
\sum_{i=1}^{T_r}
\left(
\frac{\exp(L_{b,c,i,h,w})}
{\sum_{j=1}^{T_r}\exp(L_{b,c,j,h,w})}
\right)
X^{\mathrm{ref}}_{b,c,i,h,w},
\end{equation}
where $L_{b,c,i,h,w}$ denotes the learnable logit associated with the $i$-th historical reference at location $(h,w)$ for pollutant channel $c$. The softmax normalization ensures that the contributions of different historical states are adaptively balanced at each spatial position and for each pollutant.

The final prediction is obtained by combining the backbone branch and the historical-reference branch:
\begin{equation}
\hat{\mathbf{P}}_{t}
=
\phi\left(\mathbf{F}^{\mathrm{back}} + \mathbf{R}\right),
\end{equation}

where $\mathbf{R}$ denotes the fused historical reference feature and $\phi(\cdot)$ is the final projection layer. This formulation can be interpreted as a residual temporal-guidance mechanism: the backbone branch captures large-scale meteorological and transport-driven dynamics, whereas the reference branch injects pollutant-specific persistence and recent-evolution information. By introducing spatially and channel-wise adaptive temporal weighting, the model can learn heterogeneous temporal inheritance patterns across pollutants rather than relying on a uniform persistence assumption. This is particularly important for multi-pollutant forecasting, where different variables are governed by distinct physical and chemical processes.

\section{Experiments}

\subsection{Dataset}
The input data consist of three components: pollutant concentrations, meteorological variables, and emission data. During training, meteorological inputs are derived from the European Centre for Medium-Range Weather Forecasts Reanalysis v5 (ERA5) dataset. During inference, these meteorological inputs are replaced by forecast fields from FuXi-2.0, so that the pollutant prediction model is evaluated under forecast-driven conditions. Detailed information on the dataset is provided in Appendix~A.

\subsection{Training}

\subsubsection{Loss function}

We optimize AirMPA with a hybrid weighted loss to jointly enforce reconstruction accuracy and spatial structural consistency. Specifically, the reconstruction term is formulated as a latitude-weighted Charbonnier loss to account for the non-uniform area of the global latitude--longitude grid, while an SSIM loss is introduced to preserve the spatial structure of pollutant fields. The detailed definitions of the loss terms are provided in Appendix~B.1. The overall objective is written as
\begin{equation}
\mathcal{L}
=
\alpha \mathcal{L}_{\mathrm{Lat\text{-}weighted\ Charbonnier}}
+
\beta \mathcal{L}_{\mathrm{SSIM}}.
\end{equation}
where $w_i$ denotes the latitude-dependent area weight for the $i$-th latitude band, $\phi_i$ is the corresponding latitude, and $H$ is the total number of latitude bands, $\mathcal{L}_{\text{Charbonnier}}$ is the Charbonnier reconstruction loss, $\mathcal{L}_{\text{SSIM}}$ is the structural similarity loss, and $\alpha$ and $\beta$ are stage-dependent weighting coefficients.

\subsubsection{Training stages}

AirMPA is optimized using a three-stage progressive training strategy. The first stage performs global pre-training on CAMS reanalysis data to capture large-scale spatiotemporal dependencies among meteorology, emissions, and pollutant evolution. The second stage fine-tunes the model on CAMS forecast data to adapt the output distribution to operational forecasting products. The third stage further improves the prediction of high-concentration events through extreme-oriented optimization. Detailed loss weights and optimization settings for all stages are provided in Appendix~B.2.

\subsubsection{Implementation and Training Details}

AirMPA is implemented in PyTorch and trained using the three-stage progressive strategy described above to ensure stable convergence across different pollutants and forecast horizons. The model adopts a patch size of $4$, a Swin window size of $8$, and an embedding dimension of $1024$, with $20$ stacked AirMPA blocks and $16$ attention heads in each block. In the cube embedding module, the temporal convolution uses a kernel size of $2$, and a dropout rate of $0.1$ is applied throughout the network to improve generalization. Training is performed on a single NVIDIA A100 GPU with a batch size of $2$ and requires approximately 72 hours in total. We optimize the model using AdamW ($\beta_1=0.9$, $\beta_2=0.99$, weight decay $=10^{-3}$) together with a cosine annealing learning-rate schedule.

\subsection{Evaluation}
To ensure physically consistent evaluation on a latitude–longitude grid, we adopt the latitude-weighted RMSE as the primary metric.

\begin{equation}
\text{RMSE}_{\text{lat-weight}} =
\sqrt{
\frac{\sum_{i=1}^{N_{\phi}} \sum_{j=1}^{N_{\lambda}}
\cos(\phi_i)\left(\hat{y}_{i,j}-y_{i,j}\right)^2}
{\sum_{i=1}^{N_{\phi}} \sum_{j=1}^{N_{\lambda}} \cos(\phi_i)}
},
\end{equation}

where $\hat{y}{i,j}$ and $y{i,j}$ denote the predicted and reference values at each grid point.

To evaluate extreme-event forecasting, we adopt two metrics defined at the 90th percentile threshold, namely $\text{RMSE}_{90}$ and $\text{CSI}_{90}$:
\begin{equation}
\text{RMSE}_{90} =
\sqrt{
\frac{1}{|\Omega_{90}|}
\sum_{(i,j) \in \Omega_{90}}
\left(\hat{y}_{i,j} - y_{i,j}\right)^2
},
\qquad
\text{CSI}_{90} = \frac{TP}{TP + FP + FN},
\end{equation}
where $\Omega_{90} = \{(i, j)\,|\, y_{i,j} \geq q_{0.9}\}$ is the set of grid points exceeding the 90th percentile threshold $q_{0.9}$, and $TP$, $FP$, and $FN$ denote true positives, false positives, and false negatives, respectively.

\subsection{Inference}

AirMPA performs multi-step autoregressive forecasting of atmospheric pollutant concentrations using external meteorological forecasts (Figure~\ref{fig:overview}(c)). During inference, emission priors provide background constraints on pollutant sources, FuXi-2.0 forecast fields serve as meteorological forcing, and pollutant concentration fields from CAMS FC at $t_0-12\,\mathrm{h}$ and $t_0$ are used to initialize the autoregressive process. Starting from these initial states, the model iteratively predicts pollutant fields at 12-hour intervals up to 120 hours (5 days), with each prediction conditioned on the two most recent pollutant fields together with the corresponding meteorological forecasts and emission priors. For fair comparison, AirMPA, Aurora, and CAMS use the same pollutant initialization at $t_0-12\,\mathrm{h}$ and $t_0$, so that performance differences arise from the forecasting framework itself and the meteorological inputs used thereafter.

\section{Main Results}

\subsection{Model Performance}

\begin{figure}[htbp]
  \centering
  \includegraphics[width=1\textwidth]{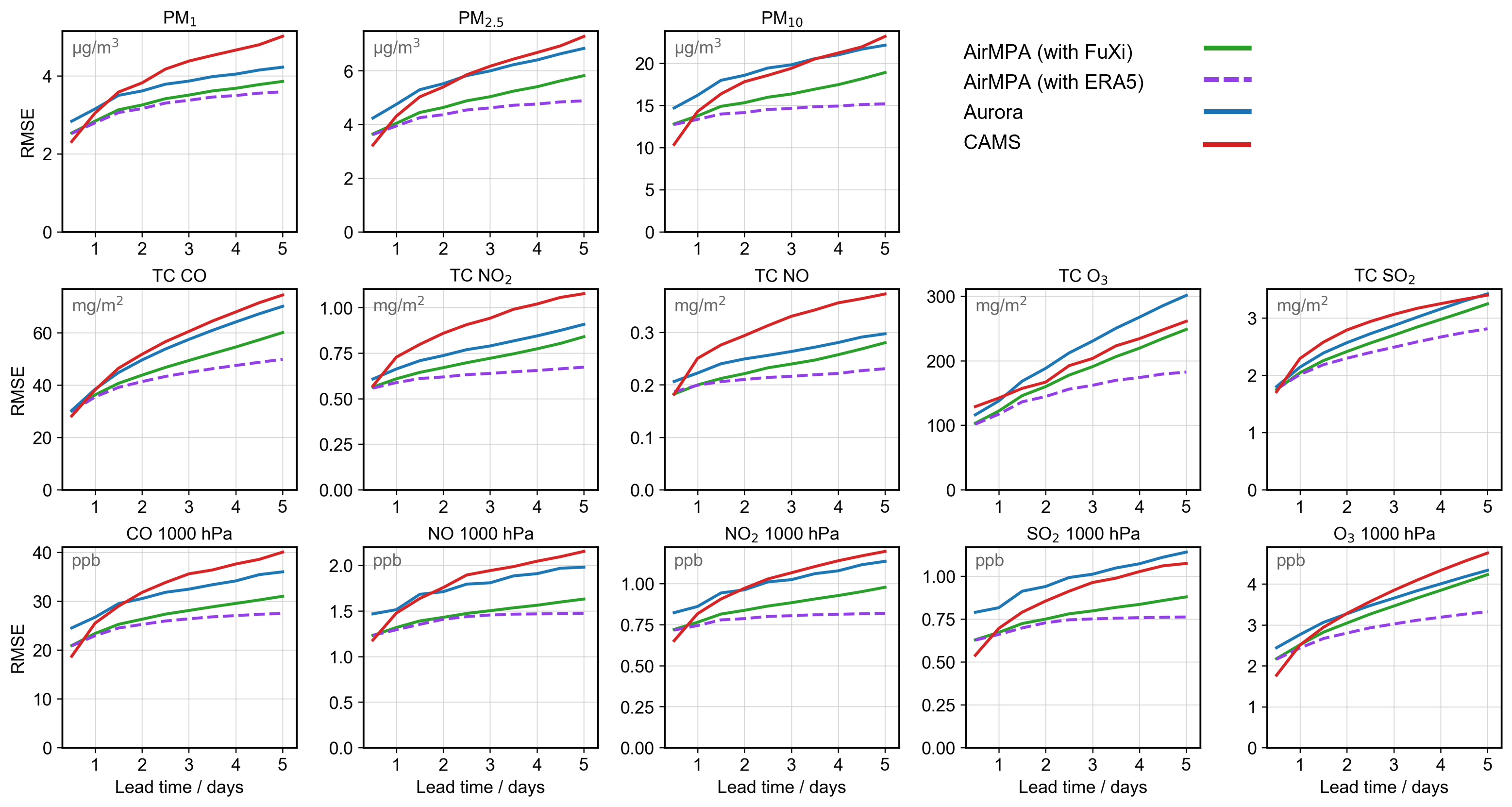}
  \caption{Comparison of forecasting performance across AirMPA (with FuXi), AirMPA (with ERA5), CAMS, and Aurora, evaluated by latitude-weighted RMSE (lower is better).}
  \label{fig:main_metrics}
\end{figure}

\textbf{Overall latitude-weighted RMSE.}
Figure~\ref{fig:main_metrics} compares the overall latitude-weighted RMSE of AirMPA (with FuXi), CAMS, and Aurora. AirMPA generally achieves lower RMSE, with larger improvements at longer lead times. We further evaluate AirMPA with ERA5 reanalysis as an offline reference. Its lower RMSE suggests that more accurate meteorological forcing could further improve pollutant forecasts; comparisons across meteorological inputs are reported in Table~\ref{tab:meteorological_forcing}. A broader comparison (Figure~\ref{fig:main_heatmap}) shows that AirMPA consistently outperforms Aurora across all 13 pollutants and surpasses CAMS on 90\% of metrics over the 5-day horizon and on 100\% beyond 48 hours. Similar improvements are observed over land-only regions, supporting the practical value of the proposed approach.

We additionally compare AirMPA with the offline
meteorology--pollution model of Fan et al.~\cite{fan2025offline}
over June--November 2022.
Across eight shared pollutant variables at lead times of
1, 3, and 5 days, AirMPA achieves lower RMSE in 23 of
24 cases, with a mean relative reduction of 16.65\%.
The only exception is total-column O$_3$ at day 5,
for which RMSE increases by 3.88\%.
Detailed results are provided in
Appendix~\ref{app:fan_baseline}
(Table~\ref{tab:fan_baseline}).
Although the evaluation period and pollutant variables
are matched, the two models use independently developed
training and preprocessing pipelines.
This comparison therefore assesses their predictive
performance rather than isolating the effect of
meteorology--pollution decoupling.

\begin{figure}[htbp]
  \centering
  \includegraphics[width=1\textwidth]{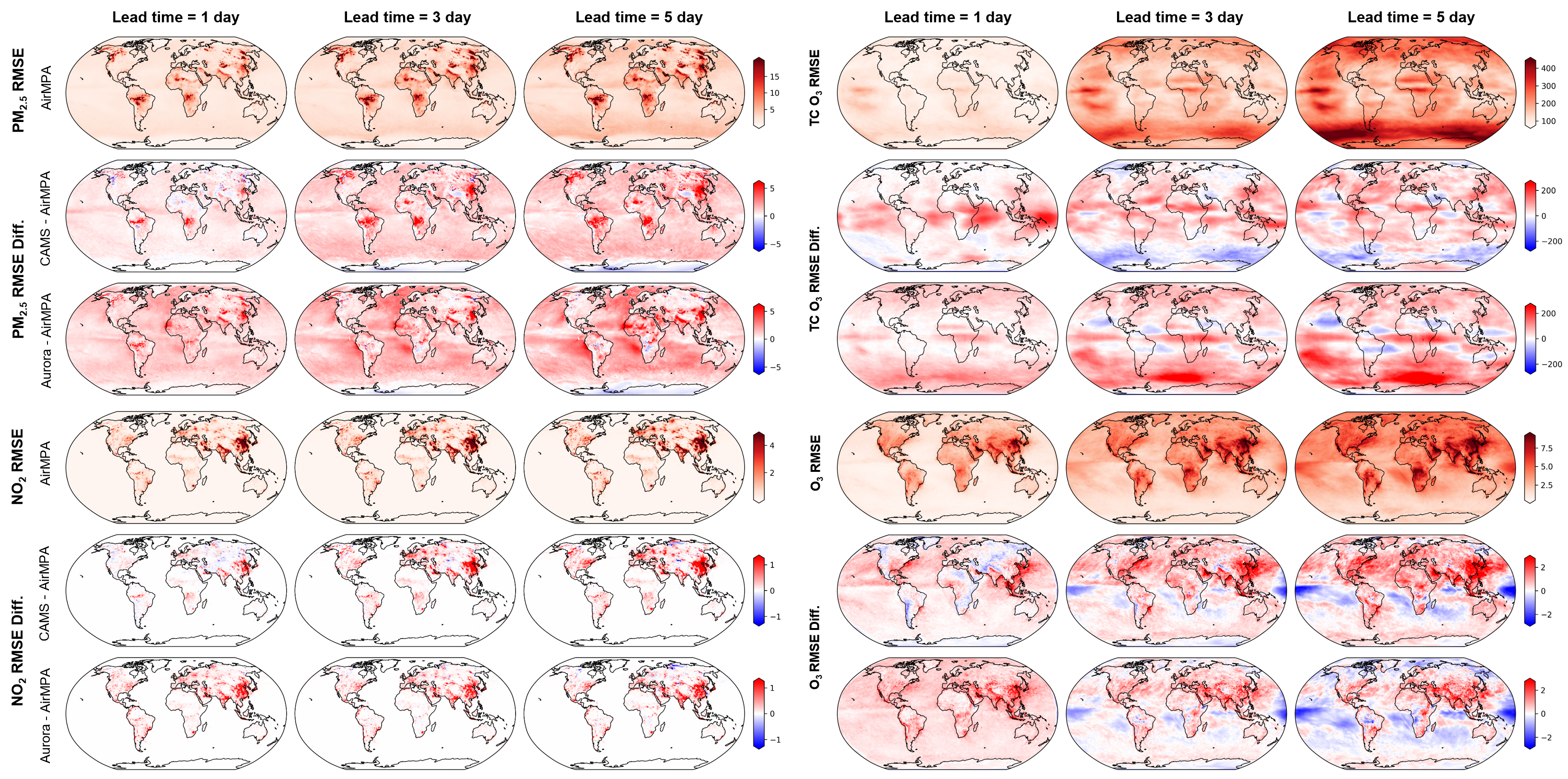}
  \caption{Spatial distribution of RMSE and RMSE differences for PM$_{2.5}$, total column O$_3$, surface NO$_2$, and surface O$_3$ over the evaluation period. In each panel, rows from top to bottom show AirMPA RMSE, RMSE difference ($\mathrm{CAMS} - \mathrm{AirMPA}$), and RMSE difference ($\mathrm{Aurora} - \mathrm{AirMPA}$). Red regions indicate better performance of AirMPA.}
  \label{fig:countourf}
\end{figure}

\textbf{Regional variation of metrics.} Figure~\ref{fig:countourf} presents the spatial variation of evaluation metrics for several key pollutants, with additional results shown in Appendix Figure~\ref{fig:all_countourf}. AirMPA consistently outperforms Aurora and CAMS in terms of RMSE across lead times, variables, and most regions. The improvements are particularly evident in densely populated and heavily polluted areas, including East Asia, the Indian subcontinent, and Central Africa. As these regions are associated with higher exposure and health risks \cite{li2023reversal,rentschler2023global}, the improved forecasting accuracy of AirMPA in these areas highlights its strong practical significance for air pollution prevention and public health protection.

\textbf{Extreme metrics.} Accurate prediction of extreme pollution events is a key objective of air quality forecasting, yet it has received limited attention in prior evaluations \cite{tao2025deep,liao2020deep}. Figure~\ref{fig:extreme_heatmap} shows that AirMPA consistently outperforms Aurora in both RMSE$_{90}$ and CSI$_{90}$ across pollutants, and achieves lower RMSE$_{90}$ than CAMS in 83\% of cases while maintaining comparable CSI$_{90}$. The advantage is particularly evident for PM$_{2.5}$, PM$_{10}$, and O$_3$ at longer lead times.

\textbf{Observation validation metrics.} We further evaluate the models using monitoring-station observations of major pollutant concentrations from China, the United States, and Europe during the evaluation period (Figure~\ref{fig:site_validation}). For PM$_{10}$ and PM$_{2.5}$, both AirMPA and Aurora achieve lower RMSE than CAMS across all three regions, with AirMPA performing best and showing slower error accumulation as forecast lead time increases. For O$_3$, AirMPA performs comparably to Aurora while still substantially outperforming CAMS. These observation-based results show that the advantages of AirMPA extend beyond gridded benchmarks to real-world monitoring data, further highlighting its practical value for operational air quality forecasting and early warning.

\subsection{Case Study}

\begin{figure}[htbp]
  \centering
  \includegraphics[width=1\textwidth]{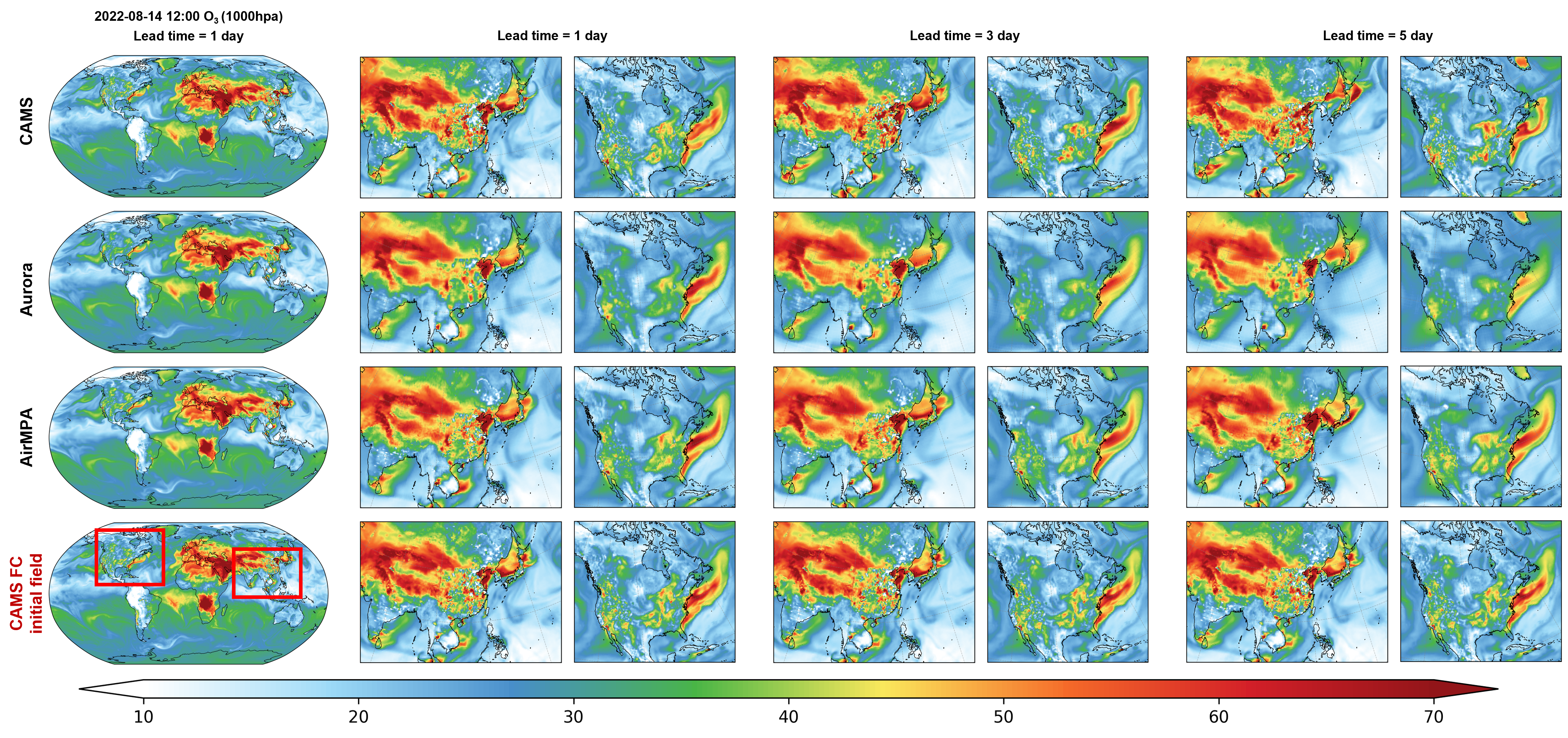}
  \caption{Comparison of surface O$_3$ forecasts at 12:00 on August 14, 2022. From top to bottom, the panels show results from CAMS, Aurora, AirMPA, and CAMS FC initial field at the valid forecast time.}
  \label{fig:case_study}
\end{figure}

Figure~\ref{fig:case_study} presents a global O$_3$ pollution event at 12:00 on August 14, 2022. Forecast results from CAMS, Aurora, and AirMPA are compared at both global and regional scales (East Asia: 75$^\circ$E--145$^\circ$E, 5$^\circ$N--55$^\circ$N; North America: 130$^\circ$W--60$^\circ$W, 20$^\circ$N--70$^\circ$N). All three models capture the overall global distribution patterns. At the regional scale, the 1-day forecasts from both AirMPA and CAMS closely match the CAMS FC reference field at the valid time. In contrast, Aurora exhibits insufficient spatial detail and underestimates extreme values over southern China and central North America, with these issues becoming more pronounced at lead times of 3 and 5 days. As a numerical model, CAMS demonstrates advantages in capturing extreme values but tends to produce spurious overestimations, such as the false high O$_3$ concentrations over Greenland at a lead time of 5 days. Although AirMPA still shows some limitations in resolving fine-scale details and tends to underestimate certain extreme values, it overall delivers the best forecasting performance over both East Asia and North America.

\subsection{Ablation Study}
\noindent\textbf{Effect of the decoupled formulation.}
We compare a decoupled Swin baseline without AirMPA-specific components against Aurora, matching its Swin architecture as closely as possible while retaining some differences in inputs and hyperparameters (Table~\ref{tab:decoupled_baseline}). The 280M-parameter baseline outperforms the 1.3B-parameter Aurora in 32 of 39 pollutant--lead-time cases, reducing RMSE by 7.20\% on average. These results suggest that the decoupled formulation itself contributes to improved forecasting performance beyond the gains from the proposed architectural components.

% ---------- Table 1: Decoupled formulation ----------
\begin{table}[t]
\centering
\caption{Decoupled Swin versus Aurora.
$\Delta$RMSE denotes the mean relative RMSE change against
Aurora; negative values indicate improvement.
Better cases count lower-RMSE results across 13 pollutants
at each lead time.}
\label{tab:decoupled_baseline}
\small
\setlength{\tabcolsep}{8pt}
\renewcommand{\arraystretch}{1.12}
\begin{tabular}{lcccc}
\toprule
Metric & 1d & 3d & 5d & Overall \\
\midrule
Better cases
& 11/13 & 11/13 & 10/13 & 32/39 \\
Mean $\Delta$RMSE
& $-5.99\%$ & $-8.17\%$ & $-7.43\%$ & $-7.20\%$ \\
\bottomrule
\end{tabular}
\end{table}

\medskip
\noindent\textbf{Sensitivity to meteorological forcing.}
We evaluate AirMPA with ERA5, FuXi, and ECMWF HRES inputs without modifying or retraining the pollutant model (Table~\ref{tab:meteorological_forcing}). FuXi and HRES yield similar performance, outperforming CAMS in 38 of 39 cases and reducing RMSE by 14.69\% and 15.26\% on average, respectively. ERA5 yields a larger reduction of 22.27\%, with the gap widening at longer lead times. These results demonstrate that AirMPA can incorporate forecasts from different meteorological systems without retraining, while the stronger performance with ERA5 suggests that more accurate meteorological forcing could further improve pollutant forecasts.

A complete pollutant-specific comparison at lead times of
1, 3, and 5 days is provided in
Appendix~\ref{app:forcing_pollutant}
(Table~\ref{tab:forcing_pollutant}).
Relative to FuXi forcing, ERA5 reduces RMSE in 37 of 39 cases,
with mean reductions of 1.19\%, 8.42\%, and 15.95\%,
respectively. HRES yields smaller mean reductions of
0.31\%, 1.45\%, and 2.35\%, with the largest benefit
for total-column SO$_2$.

% ---------- Table 2: Meteorological forcing ----------
\begin{table}[htbp]
\centering
\caption{AirMPA performance under different meteorological
inputs. Values are mean relative RMSE changes against CAMS
across 13 pollutants; negative values indicate improvement.
Overall averages the 39 pollutant--lead-time cases.
ERA5 provides an offline reanalysis reference.}
\label{tab:meteorological_forcing}
\small
\setlength{\tabcolsep}{8pt}
\renewcommand{\arraystretch}{1.12}
\begin{tabular}{lcccc}
\toprule
Input & 1d & 3d & 5d & Overall \\
\midrule
ERA5
& $-10.72\%$ & $-24.89\%$ & $-31.21\%$ & $-22.27\%$ \\
FuXi
& $-8.63\%$ & $-17.65\%$ & $-17.80\%$ & $-14.69\%$ \\
HRES
& $-8.81\%$ & $-18.15\%$ & $-18.80\%$ & $-15.26\%$ \\
\bottomrule
\end{tabular}
\end{table}

\medskip
\noindent\textbf{Contributions of architectural components.}
To validate the effectiveness of the proposed architectural components, we conduct ablation studies (Table~\ref{tab:ablation}) and evaluate different model variants using the latitude-weighted RMSE over a 5-day forecasting horizon (Table~\ref{tab:ablation_results}). The results show that AirMPA achieves the best or tied-best performance across all reported variables and lead times, while different components contribute to different pollutant types in distinct ways. For emission-sensitive particulate pollutants such as PM$_{2.5}$, incorporating emission inputs consistently improves forecasting performance across the reported lead times, highlighting the role of emission priors in capturing the temporal evolution of local pollution. In contrast, near-surface O$_3$ and total-column NO$_2$ exhibit lower sensitivity to emission inputs; for long-range forecasts, adding emission inputs may even increase RMSE. The zonal release attention mechanism provides limited benefit for local pollutants such as PM$_{2.5}$, but improves the prediction of total-column NO$_2$ and O$_3$, consistent with its intended role in representing large-scale transport continuity.

Overall, introducing emission features reduces the average RMSE by approximately 0.8--2.1 units. Cube embedding and zonal release attention yield moderate gains, whereas the model with dual-branch fusion delivers the second-best performance in nearly all experiments. Relative to the original Swin baseline, the variant incorporating emission inputs and dual-branch fusion lowers the average RMSE by 7.9\% for 1-day forecasts and 10.1\% for 5-day forecasts, supporting its contribution to the performance of the full model.

% ---------- Table 3: Architectural components ----------
\begin{table*}[ht]
\centering
\caption{Ablation results of key architectural components
in AirMPA (lower is better).
PM$_{2.5}$ is in $\mathrm{\mu g}/\mathrm{m}^3$,
O$_3$ is in ppb, and TC NO$_2$ is in $\mathrm{mg}/\mathrm{m}^2$.
Best values are shown in bold, and second-best values
are underlined.
$^{*}$ indicates models built on the version with emission input.}
\label{tab:ablation_results}
\small
\setlength{\tabcolsep}{4pt}
\renewcommand{\arraystretch}{1.12}
\begin{tabular}{>{\centering\arraybackslash}m{3.8cm}cccccccc}
\toprule
\multirow{2}{*}{\textbf{Model}}
& \multicolumn{2}{c}{\textbf{PM$_{2.5}$}$\downarrow$}
& \multicolumn{2}{c}{\textbf{O$_3$}$\downarrow$}
& \multicolumn{2}{c}{\textbf{TC NO$_2$}$\downarrow$}
& \multicolumn{2}{c}{\textbf{Average RMSE}$\downarrow$} \\
\cmidrule(lr){2-3}
\cmidrule(lr){4-5}
\cmidrule(lr){6-7}
\cmidrule(lr){8-9}
& \textbf{1d} & \textbf{5d}
& \textbf{1d} & \textbf{5d}
& \textbf{1d} & \textbf{5d}
& \textbf{1d} & \textbf{5d} \\
\midrule
Swin Transformer (Baseline)
& 4.31 & 6.08
& 2.595 & 4.435
& 0.651 & 0.870
& 17.68 & 32.20 \\

w/ Emission Input
& 4.29 & 6.03
& 2.590 & 4.630
& 0.648 & 0.872
& 16.84 & 30.08 \\

w/ Cube Embedding$^{*}$
& 4.25 & 6.13
& 2.558 & 4.425
& \underline{0.619} & \underline{0.845}
& 16.76 & 30.00 \\

w/ Zonal Release Attention$^{*}$
& 4.28 & 6.03
& 2.543 & 4.420
& 0.627 & 0.864
& 16.80 & 30.05 \\

w/ Dual-branch Fusion$^{*}$
& \underline{4.12} & \underline{5.79}
& \underline{2.540} & \underline{4.361}
& 0.622 & \textbf{0.826}
& \underline{16.28} & \underline{28.95} \\

AirMPA
& \textbf{4.02} & \textbf{5.72}
& \textbf{2.505} & \textbf{4.345}
& \textbf{0.610} & \textbf{0.826}
& \textbf{16.05} & \textbf{28.86} \\
\bottomrule
\end{tabular}
\end{table*}

\section{Conclusions}
We propose AirMPA, a decoupled autoregressive framework for global air pollutant forecasting that separates meteorological forecasting from pollutant prediction. By reformulating air quality prediction as a meteorology-driven downstream task, AirMPA establishes a flexible and scalable paradigm that can directly interface with external meteorological foundation models and can readily benefit from improvements in meteorological forecasting under the proposed decoupled formulation. Built upon an improved Swin-Transformer architecture, the model integrates zonal release attention and dual-branch fusion to effectively capture the continuity of global pollutant transport and the heterogeneous dynamics of different pollutants, resulting in consistently strong forecasting performance across variables and lead times. AirMPA provides a practical and extensible pathway for bringing meteorological foundation models into air quality forecasting, and lays a foundation for future fully AI-driven global atmospheric composition prediction systems.

\clearpage

\bibliographystyle{unsrt}
\bibliography{references}

\clearpage

\appendix

\section*{Appendix}

\section{Dataset}

\subsection{Air pollutant concentrations}
The pollutant data used in this study comprise two CAMS
products: the CAMS global reanalysis (EAC4; hereafter RA)
\cite{inness2019cams}
\footnote{\url{https://ads.atmosphere.copernicus.eu/datasets/cams-global-reanalysis-eac4}}
and the CAMS global atmospheric composition forecasts
(hereafter FC)
\footnote{\url{https://ads.atmosphere.copernicus.eu/datasets/cams-global-atmospheric-composition-forecasts}}.
The RA dataset provides a long-term, temporally continuous
record for model pre-training, while the FC dataset is used
for model adaptation and evaluation.

\subsection{ERA5}
This study uses the European Centre for Medium-Range Weather Forecasts Reanalysis v5 (ERA5) reanalysis dataset \cite{hersbach2020era5} provided by the European Centre for Medium-Range Weather Forecasts (ECMWF). To ensure consistency with the FuXi meteorological forecast data, the precipitation variable is processed as a 12-hour accumulated precipitation field during preprocessing.

\subsection{FuXi-2.0}
The FuXi-2.0 model provides global meteorological forecasts at a $0.25^\circ \times 1\text{-h}$ resolution \cite{zhong2024FuXi2}. The dataset encompasses 5 upper-air variables (geopotential height, temperature, humidity, and $u/v$ wind components) across 13 standard pressure levels (from 1000 to 50~hPa), alongside 23 surface variables covering critical atmospheric and oceanic conditions (e.g., 2-m temperature, sea level pressure, and wave parameters). For operational consistency and real-time efficiency, this study utilizes the 6-hour forecast version of FuXi-2.0. During inference, these outputs are resampled to align with the spatiotemporal resolution of the CAMS and ERA5 dataset.

\subsection{Emission inventories}
We adopt weekday, weekend, and hourly allocation factors
from Wang et al.~\cite{wang2014source}, following a temporal
disaggregation approach similar to CAMS-GLOB-TEMPO \footnote{\url{https://ads.atmosphere.copernicus.eu/datasets/cams-global-emission-inventories}}.
The allocation is performed in local time at each grid cell,
with separate normalization at the daily and hourly levels
to preserve monthly emission totals.

Let $E_{g,s,m}$ denote the monthly emission total for grid cell
$g$, emission species $s$, and month $m$ after conservative
remapping. For each local calendar day $d$ in month $m$, we
assign a factor $a_d$ according to whether it is a weekday
or a weekend. The daily emission total is calculated as
\begin{equation}
E_{g,s,d}
=
E_{g,s,m}
\frac{a_d}{\sum_{d' \in \mathcal{D}_m} a_{d'}},
\end{equation}
where $\mathcal{D}_m$ is the set of local calendar days in
month $m$. This normalization accounts for the actual
numbers of weekdays and weekend days in each month.

The hourly allocation profile is constructed by combining
source-specific activity profiles:
\begin{equation}
b_{g,s,d,h}
=
\sum_{k \in \mathcal{K}}
w_{g,s,m,k}\,p_{k,d,h},
\end{equation}
where $\mathcal{K}$ denotes the source categories,
$w_{g,s,k}$ is the contribution weight of source $k$, and
$p_{k,d,h}$ is its activity factor at local hour
$h \in \{0,\ldots,23\}$ on day $d$.
Hourly emissions are then obtained as
\begin{equation}
E_{g,s,d,h}
=
E_{g,s,d}
\frac{b_{g,s,d,h}}
{\sum_{h'=0}^{23} b_{g,s,d,h'}}.
\end{equation}
Consequently,
\begin{equation}
\sum_{d \in \mathcal{D}_m}
\sum_{h=0}^{23}
E_{g,s,d,h}
=
E_{g,s,m}.
\end{equation}

Local dates and hours are determined from the UTC timestamp
using the time-zone offset assigned to each grid cell:
\begin{equation}
t_{\mathrm{local}}(g)
=
t_{\mathrm{UTC}} + \Delta_g,
\end{equation}
where $\Delta_g$ is the local UTC offset in hours.
The local calendar date determines the weekday or weekend
factor, while the local hour determines the hourly activity
factor. After temporal allocation, the hourly emissions are
indexed in UTC and accumulated over the 12-hour interval
ending at each model input time.

\subsection{Data preprocess}
To address unit inconsistencies among variables in the original CAMS outputs and enhance the physical interpretability of the model inputs, all variables were converted into commonly used units in atmospheric modeling. Specifically, particulate matter concentrations ($\text{PM}_{1}$, $\text{PM}_{2.5}$, and $\text{PM}_{10}$) were converted to $\mu\mathrm{g/m^3}$, while total column concentrations were expressed in $\mathrm{mg/m^2}$. For gas concentrations at the 1000~hPa level, mass mixing ratios ($\mathrm{kg/kg}$) were converted to volume mixing ratios ($\mathrm{ppbv}$) using the ideal gas law, as formulated in Eq.~\ref{eq:mixing_ratio}.

\begin{equation}
C_{\text{ppbv}} = C_{\text{kg/kg}} \times \frac{M_{\text{air}}}{M_{\text{gas}}} \times 10^9
\label{eq:mixing_ratio}
\end{equation}

where $C_{\text{ppbv}}$ denotes the volume mixing ratio, $C_{\text{kg/kg}}$ represents the mass mixing ratio, and $M_{\text{air}}$ (28.97) and $M_{\text{gas}}$ denote the mean molecular weights of air and the corresponding gas species, respectively.

To balance the training data and incorporate multi-source information, a total of 78 variables, including emissions, meteorological variables, and pollutant concentrations, were used as model inputs. All datasets were resampled to match the spatial resolution of the CAMS forecast data ($0.4^\circ$). Due to the highly skewed distribution of emission variables, a log transformation was applied before model input to stabilize the data distribution:

\begin{equation}
\text{log1p}(x) = \ln(1 + x)
\end{equation}

Subsequently, all variables were standardized using z-score normalization. In addition, four temporal features—day of year and hour of day encoded using sinusoidal functions—were introduced and broadcast across the spatial domain:

\begin{equation}
h_{\sin} = \sin \left( \frac{2\pi h}{24} \right), \quad
h_{\cos} = \cos \left( \frac{2\pi h}{24} \right)
\end{equation}

\begin{equation}
d_{\sin} = \sin \left( \frac{2\pi d}{T} \right), \quad
d_{\cos} = \cos \left( \frac{2\pi d}{T} \right)
\end{equation}

where $h$ denotes the hour of the day, $d$ denotes the day of the year, and $T$ represents the total number of days in a year.

\begin{table*}[htbp]
\centering
\caption{Model Input Data}
\label{tab:input_data}
\footnotesize
\setlength{\tabcolsep}{0.5pt}
\renewcommand{\arraystretch}{1.2}

\begin{tabularx}{\textwidth}{
>{\centering\arraybackslash}p{1.55cm}
>{\centering\arraybackslash}p{1.85cm}
>{\raggedright\arraybackslash}X
>{\centering\arraybackslash}p{1.8cm}
>{\centering\arraybackslash}p{1.35cm}
>{\centering\arraybackslash}p{1.75cm}
}
\toprule
\textbf{Data Type} & \textbf{Source} & \textbf{Variables} & \textbf{Spatial Res.} & \textbf{Temporal Res.} & \textbf{Period} \\
\midrule

Pollutants
& CAMS-RA
& \makecell[l]{pm1, pm2p5, pm10, tcco, tcno2, tc\_no,\\
gtco3, tcso2, co\_1000, no\_1000,\\
no2\_1000, so2\_1000, go3\_1000}
& $0.75^\circ\to0.4^\circ$
& 12 h
& 2010.01--2021.12 \\
\cmidrule(lr){2-6}

& CAMS-FC
& Same as above
& $0.4^\circ$
& 12 h
& 2017.10--2022.05 \\
\midrule

Meteorology
& ERA5
& \makecell[l]{U10m, v10m, d2m, t2m, msl, tcc, tp;\\
z, u, v, t (at 13 levels: 50, 100, 150, 200, 250, \\
300, 400, 500, 600, 700, 850, 925, 1000 hPa)}
& $0.25^\circ\to0.4^\circ$
& 12 h
& 2010.01--2023.12 \\
\cmidrule(lr){2-6}

& FuXi
& \makecell[l]{U10m, v10m, d2m, t2m, msl, tcc, tp;\\
z, u, v, t (at 13 levels)}
& $0.25^\circ\to0.4^\circ$
& 12 h
& 2022.06--2022.11 \\
\midrule

Emissions
& CAMS emission
& \makecell[l]{ammonia, black carbon, carbon monoxide,\\
nitrogen oxides, NMVOCs, sulfur dioxide}
& $0.1^\circ\to0.4^\circ$
& 12 h
& 2010.01--2023.12 \\
\bottomrule
\end{tabularx}
\end{table*}

\subsection{Dataset partitioning}
Considering the data quality and temporal stability of CAMS forecast products, and following the evaluation protocol of Aurora \cite{bodnar2025foundation}, we adopt a three-stage training strategy. Since the CAMS FC dataset provides a relatively limited number of samples and has only offered stable outputs for particulate matter, total column concentrations, and near-surface gas concentrations since October 2017, we first pre-train the model on the CAMS RA dataset from January 2010 to December 2021. This stage enables the model to learn the large-scale spatiotemporal distribution and evolution patterns of atmospheric pollutants under the influence of meteorological conditions and emission fields. The pre-trained model is then fine-tuned on the CAMS FC dataset from October 2017 to May 2022, so that the model output distribution can better align with the FC products used for downstream evaluation and comparison.

For model development, data from January 2023 to December 2023 are used as the validation set, while an independent test period from June 2022 to November 2022 is adopted for final evaluation (Table~\ref{tab:data_split}). Early stopping with a patience of 20 epochs is applied in Stages 2 and 3 based on the validation RMSE.

\begin{table}[htbp]
\centering
\caption{Dataset partitioning and time ranges for different training stages.}
\label{tab:data_split}
\footnotesize
\renewcommand{\arraystretch}{1.2}
\setlength{\tabcolsep}{4pt}
\begin{tabular}{llll}
\toprule
\textbf{Stage / Split} & \textbf{Dataset} & \textbf{Purpose} & \textbf{Period} \\
\midrule
Stage 1 & CAMS-RA & Global pre-training & 2010.01--2021.12 \\
Stage 2 & CAMS-FC & Operational fine-tuning & 2017.10--2022.05 \\
Test & CAMS-FC & Final evaluation & 2022.06--2022.11 \\
\bottomrule
\end{tabular}
\end{table}

\section{Experimental Details}

\subsection{Loss Function}

This appendix provides the detailed definitions of the two loss terms used in AirMPA, namely the latitude-weighted Charbonnier loss and the structural similarity (SSIM) loss.

\subsubsection{Latitude-weighted Charbonnier loss}

To account for the non-uniform area represented by the global latitude--longitude grid, we use a latitude-weighted Charbonnier loss:
\begin{equation}
\mathcal{L}_{\text{Lat-weighted Charbonnier}}
=
\sum_{i=1}^{H}\sum_{j=1}^{W}
\frac{\cos(\phi_i)}{\sum_{i'=1}^{H}\cos(\phi_{i'})}
\sqrt{(\hat{y}_{i,j}-y_{i,j})^2+\epsilon^2},
\end{equation}
where $\hat{y}_{i,j}$ and $y_{i,j}$ denote the predicted and reference pollutant values at grid cell $(i,j)$, respectively, $\phi_i$ is the latitude of the $i$-th row, $H$ and $W$ are the numbers of latitude and longitude grid cells, and $\epsilon$ is a small constant for numerical stability.

\subsubsection{Structural similarity loss}

To preserve the spatial structure of pollutant fields, we further introduce the SSIM loss. The structural similarity index is defined as
\begin{equation}
\mathrm{SSIM}(\hat{y},y)
=
\frac{(2\mu_{\hat{y}}\mu_y+C_1)(2\sigma_{\hat{y}y}+C_2)}
{(\mu_{\hat{y}}^2+\mu_y^2+C_1)(\sigma_{\hat{y}}^2+\sigma_y^2+C_2)},
\end{equation}
where $\mu_{\hat{y}}$ and $\mu_y$ are the mean values of the predicted and reference fields, $\sigma_{\hat{y}}^2$ and $\sigma_y^2$ are the corresponding variances, $\sigma_{\hat{y}y}$ is the covariance, and $C_1$ and $C_2$ are constants used to stabilize the computation.

The corresponding SSIM loss is defined as
\begin{equation}
\mathcal{L}_{\text{SSIM}}
=
1-\mathrm{SSIM}(\hat{y},y).
\end{equation}

\subsection{Training stages}

AirMPA is optimized using a three-stage progressive training strategy, including global pre-training, operational fine-tuning, and extreme-event optimization. The detailed settings of each stage are summarized in Table~\ref{tab:training_stages}. In the third stage, the dataset is further partitioned using pollutant-specific extreme-value thresholds, and higher loss weights are assigned to these critical samples to improve the reconstruction of high-concentration events. The manually specified extreme thresholds used for Stage 3 training and the 90th percentile thresholds used for test-period evaluation are listed in Table~\ref{tab:thresholds}.

\begin{table}[htbp]
\centering
\caption{Training objectives and optimization settings for the three training stages of AirMPA.}
\label{tab:training_stages}

\footnotesize
\setlength{\tabcolsep}{4pt}
\renewcommand{\arraystretch}{1.2}

\begin{tabular*}{\textwidth}{@{\extracolsep{\fill}}p{1.4cm}p{3.2cm}p{2.8cm}p{4.0cm}}
\toprule
\textbf{Stage} & \textbf{Objective} & \textbf{Learning rate} & \textbf{Loss settings} \\
\midrule
Stage 1 &
Global pre-training on CAMS reanalysis &
$2\times10^{-4} \rightarrow 10^{-6}$ &
Hybrid loss: $\alpha=1.0$, $\beta=0.25$, $\epsilon=10^{-3}$ \\
\addlinespace
Stage 2 &
Operational fine-tuning on CAMS forecast data &
$5\times10^{-5} \rightarrow 10^{-6}$ &
Hybrid loss: $\alpha=1.0$, $\beta=0.75$, $\epsilon=10^{-3}$ \\
\addlinespace
Stage 3 &
Extreme-event optimization &
continued from Stage 2 &
Weighted RMSE loss: overall RMSE $=0.5$, extreme-event RMSE $=0.5$ \\
\bottomrule
\end{tabular*}
\end{table}

\begin{table*}[bthp]
\centering
\caption{Extreme thresholds and 90th percentile values for different variables.}
\label{tab:thresholds}
\footnotesize
\setlength{\tabcolsep}{4pt}
\renewcommand{\arraystretch}{1.2}

\begin{tabularx}{\textwidth}{l *{8}{>{\centering\arraybackslash}X}}
\toprule
\textbf{Variable} 
& \textbf{PM$_{1}$} & \textbf{PM$_{2.5}$} & \textbf{PM$_{10}$} 
& \textbf{TC CO} & \textbf{TC NO} & \textbf{TC NO$_2$} & \textbf{TC SO$_2$} & \textbf{TC O$_3$} \\
\midrule

Training extreme threshold 
& 8.0 & 25.0 & 50.0 
& 900 & 1.5 & 3.5 & 3.0 & 7500 \\

Threshold (90\%) 
& 5.2323 & 20.8496 & 42.1983 
& 811.4433 & 1.4372 & 3.2538 & 1.7426 & 7227.6772 \\

\midrule

\textbf{Variable (at 1000 hpa)} 
& \textbf{CO} & \textbf{NO} & \textbf{NO$_2$} & \textbf{SO$_2$} & \textbf{O$_3$}
& \multicolumn{3}{c}{} \\
\midrule

Training extreme threshold 
& 135 & 0.05 & 0.8 & 0.5 & 45 
& \multicolumn{3}{c}{} \\

Threshold (90\%) 
& 110.1245 & 0.0265 & 0.4042 & 0.2312 & 38.8720 
& \multicolumn{3}{c}{} \\

\bottomrule
\end{tabularx}
\vspace{4mm}
\end{table*}

\subsection{Ablation setup}

\begin{table}[H]
\centering
\caption{Ablation study of key architectural components in AirMPA.}
\label{tab:ablation}
\footnotesize
\setlength{\tabcolsep}{4pt}
\renewcommand{\arraystretch}{1.0}
\begin{tabular}{lcccc}
\toprule
\textbf{Model} & \textbf{Zonal} & \textbf{Cube} & \textbf{Dual-Branch} & \textbf{Emission} \\
 & \textbf{Release Attention} & \textbf{Embedding} & \textbf{Fusion} &  \textbf{Input}\\
\midrule
Swin-Transformer (backbone)& $\times$ & $\times$ & $\times$ & $\times$ \\
w/ Emission & $\times$ & $\times$ & $\times$ & \checkmark \\
w/ Zonal Release Attention & \checkmark & $\times$ & $\times$ & \checkmark \\
w/ Cube Embedding & $\times$ & \checkmark & $\times$ & \checkmark \\
w/ Dual-Branch Fusion & $\times$ & $\times$ & \checkmark & \checkmark \\
AirMPA & \checkmark & \checkmark & \checkmark & \checkmark \\
\bottomrule
\end{tabular}
\end{table}

\subsection{Compute resources}
Training AirMPA required approximately 72 GPU-hours on one NVIDIA A100 GPU with 80 GB memory. Each ablation variant required approximately 72 GPU-hours. Inference over the six-month test period required approximately 1.5 GPU-hours. The processed datasets occupied approximately 2 TB of storage. Preliminary experiments are not included in these estimates.

\section{Limitations}
Despite the strong performance, several limitations remain and point to directions for future work.

First, the model is trained on CAMS atmospheric composition reanalysis and forecast datasets, where initial conditions are derived from a combination of previous forecasts and satellite observations. The limited use of surface monitoring data may reduce consistency with real-world observations. Incorporating additional observational constraints and physical conservation laws could further improve short-term stability and physical realism. 

Second, real-world emissions are highly dynamic and influenced by factors such as holidays, industrial activities, and traffic regulations. In the current framework, emission inputs are primarily based on hourly empirical factors, which are insufficient to capture rapid or abrupt emission changes. This may introduce systematic biases under scenarios with sudden emission variations, such as large-scale events or emergency control measures. 

Third, due to the resolution of the training data, the model operates at a temporal resolution of 12 hours and a spatial resolution of $0.4^\circ$. Such resolution may be inadequate for resolving localized pollution events at finer regional scales. In addition, under the current 12-hour forecasting interval, the dual-branch fusion module only learns to adaptively weight the two preceding pollutant concentration fields, which may not fully exploit its potential for modeling more detailed temporal dependencies. Future work should extend the framework to finer temporal resolutions to further evaluate and enhance the effectiveness of this component.

\section{Broader impacts}
The development of AirMPA provides a new technical paradigm for atmospheric environmental forecasting, with important implications for both scientific research and operational air quality management. By decoupling meteorological forecasting from pollutant prediction, our framework shows that high-quality outputs from meteorological foundation models can be effectively transferred to downstream air quality forecasting tasks. This modular design improves forecasting flexibility and accuracy, and provides a scalable technical basis for future AI-driven atmospheric composition prediction systems.

From a public health perspective, atmospheric pollution remains one of the most pervasive global environmental risks. The ability of AirMPA to provide more accurate global forecasts, especially for extreme pollution events, can support earlier warnings and more timely interventions. This may help reduce human exposure to harmful pollutants and improve decision-making for environmental protection and public health management.

More broadly, the significance of this work lies in improving the predictive capability of global air quality forecasting and in providing a technical foundation for future full-chain forecasting systems. In particular, the proposed framework offers a feasible pathway toward an integrated system that combines data assimilation, ensemble meteorological prediction, and downstream pollutant forecasting. Such a system could further enhance the timeliness, robustness, and operational value of air pollution early warning.

At the same time, several limitations should be noted. Since the current model is trained primarily on reanalysis and forecast products, it may inherit systematic biases between these data sources and real-world observations, which could reduce its effectiveness in practical applications. In addition, like other data-driven forecasting models, AirMPA still requires further validation under unprecedented extreme events or rapidly changing emission conditions. Future work should strengthen the integration of ground-based observations, uncertainty quantification, and physical consistency constraints, so as to improve reliability and interpretability in real-world decision-making.

\section{Code and data availability}

The ERA5 and CAMS products used in this study were obtained from the Copernicus Climate Data Store (CDS; \url{https://cds.climate.copernicus.eu/}) and the Copernicus Atmosphere Data Store (ADS; \url{https://ads.atmosphere.copernicus.eu/datasets}), respectively, under the Copernicus CC-BY licence. In accordance with the Copernicus terms, the derived products contain modified Copernicus Climate Change Service and Copernicus Atmosphere Monitoring Service information, and neither the European Commission nor ECMWF is responsible for any use made of the information or data contained herein. The ERA5 reanalysis data were downloaded, subsetted, and preprocessed for use in the AirMPA model. The CAMS reanalysis and forecast data, together with anthropogenic emission inventories, were used for model comparison and inference experiments.

The FuXi-2.0 forecast fields were generated from operational FuXi-2.0 outputs and used for academic research with authorization from the model developers. The corresponding source code and inference scripts are available from the official repository (\url{https://github.com/tpys/FuXi2/}), and public FuXi assets were used in accordance with their stated licence terms. The Aurora code, checkpoints, and inference scripts were obtained from the official Microsoft example (\url{https://microsoft.github.io/aurora/example_cams.html}) and used under the MIT license. For fair comparison, the same CAMS dataset covering June 2022 to November 2022 was used for Aurora inference experiments.

The AirMPA implementation is publicly available at
\url{https://github.com/yhWang0712/AirMPA}.

\section{Additional results}

\subsection{Comparison with the offline model of Fan et al.}
\label{app:fan_baseline}
\begin{table}[htbp]
\centering
\caption{Relative RMSE reduction (\%) of AirMPA compared
with the offline meteorology--pollution model of
Fan et al.~\cite{fan2025offline} over June--November 2022.
Positive values (in bold) indicate lower RMSE for AirMPA,
while negative values indicate degradation.
The comparison covers eight shared pollutant variables
at lead times of 1, 3, and 5 days.}
\label{tab:fan_baseline}
\small
\setlength{\tabcolsep}{12pt}
\renewcommand{\arraystretch}{1.12}
\begin{tabular}{lrrr}
\toprule
\textbf{Variable} & \textbf{Day 1} & \textbf{Day 3} & \textbf{Day 5} \\
\midrule
PM$_1$     & \textbf{19.36} & \textbf{13.88} & \textbf{11.79} \\
PM$_{2.5}$ & \textbf{21.89} & \textbf{15.17} & \textbf{8.10}  \\
PM$_{10}$  & \textbf{33.46} & \textbf{30.39} & \textbf{24.90} \\
TC CO      & \textbf{15.38} & \textbf{13.58} & \textbf{18.14} \\
TC NO$_2$  & \textbf{20.18} & \textbf{13.84} & \textbf{8.10}  \\
TC NO      & \textbf{29.33} & \textbf{24.76} & \textbf{18.13} \\
TC O$_3$   & \textbf{25.79} & \textbf{8.36}  & -3.88         \\
TC SO$_2$  & \textbf{20.16} & \textbf{7.91}  & \textbf{0.95}  \\
\bottomrule
\end{tabular}
\end{table}

\subsection{Pollutant-specific sensitivity to meteorological forcing}
\label{app:forcing_pollutant}

We compare ERA5, FuXi, and ECMWF HRES as meteorological
inputs to the same AirMPA pollutant model, without modifying
or retraining its parameters.
Table~\ref{tab:forcing_pollutant} reports results for all
13 pollutant variables at lead times of 1, 3, and 5 days.
For each variable and lead time, we compute the relative
RMSE reduction with respect to FuXi forcing as
\begin{equation}
    R_X =
    \frac{\mathrm{RMSE}_{\mathrm{FuXi}} -
          \mathrm{RMSE}_{X}}
         {\mathrm{RMSE}_{\mathrm{FuXi}}}
    \times 100\%,
    \qquad X \in \{\mathrm{ERA5}, \mathrm{HRES}\}.
\end{equation}
Positive values indicate lower RMSE than with FuXi.
We average these percentage reductions across variables,
rather than averaging absolute RMSE values with different units.

\paragraph{ERA5 versus FuXi.}
ERA5 yields lower RMSE than FuXi in 37 of the 39
pollutant--lead-time cases.
The mean reduction increases from 1.19\% at day 1
to 8.42\% at day 3 and 15.95\% at day 5.
The two exceptions are NO and SO$_2$ at 1000 hPa
at day 1, for which RMSE increases by 5.80\% and
4.64\%, respectively.
At day 5, the largest reductions occur for total-column
O$_3$ (26.62\%) and O$_3$ at 1000 hPa (21.44\%).
The reductions for PM$_1$, PM$_{2.5}$, and PM$_{10}$
are 6.94\%, 16.12\%, and 19.66\%, respectively.
Total-column CO, NO$_2$, and NO improve by
17.06\%, 19.92\%, and 17.69\%, while their
1000-hPa counterparts improve by 11.28\%, 16.31\%,
and 8.12\%.
Total-column SO$_2$ and SO$_2$ at 1000 hPa improve
by 13.39\% and 12.77\%, respectively.

\paragraph{HRES versus FuXi.}
HRES produces smaller mean changes relative to FuXi,
reducing RMSE by 0.31\%, 1.45\%, and 2.35\%
at days 1, 3, and 5, respectively.
It yields lower RMSE in 32 of the 39 cases.
The largest reduction is for total-column SO$_2$,
reaching 17.80\% at day 5.
For this variable, HRES also yields lower RMSE
than ERA5 at all three reported lead times.
In contrast, total-column CO and O$_3$ have slightly
higher RMSE with HRES than with FuXi at all three
lead times, with increases of 0.25--1.57\% and
0.32--1.02\%, respectively.
For the remaining variables, the day-5 reductions
range from 0.24\% for CO at 1000 hPa to 4.46\%
for total-column NO.

These results show that the effect of changing
meteorological forcing varies across pollutants
and forecast horizons.
The widening gap between ERA5 and FuXi at longer
lead times suggests increasing sensitivity to the
meteorological input during autoregressive forecasting.
ERA5 provides a retrospective reanalysis reference,
so its advantage should not be interpreted as
an operationally available forecast gain.

\begin{table}[htbp]
\centering
\caption{Pollutant-specific comparison of meteorological
forcing at lead times of 1, 3, and 5 days.
Values are relative RMSE reductions (\%) compared with
FuXi forcing. Positive values (in bold) indicate lower
RMSE, while negative values indicate higher RMSE.
TC denotes total column.
The mean is the unweighted average across the 13 variables.
ERA5 provides an offline reanalysis reference.}
\label{tab:forcing_pollutant}
\small
\setlength{\tabcolsep}{5pt}
\renewcommand{\arraystretch}{1.08}
\begin{tabular}{lrrrrrr}
\toprule
& \multicolumn{2}{c}{Day 1}
& \multicolumn{2}{c}{Day 3}
& \multicolumn{2}{c}{Day 5} \\
\cmidrule(lr){2-3}
\cmidrule(lr){4-5}
\cmidrule(lr){6-7}
Variable & ERA5 & HRES & ERA5 & HRES & ERA5 & HRES \\
\midrule
PM$_1$
& \textbf{1.53} & -0.08
& \textbf{3.79} & \textbf{0.17}
& \textbf{6.94} & \textbf{0.38} \\

PM$_{2.5}$
& \textbf{2.46} & \textbf{0.13}
& \textbf{8.29} & \textbf{0.48}
& \textbf{16.12} & \textbf{0.82} \\

PM$_{10}$
& \textbf{2.99} & \textbf{0.32}
& \textbf{10.44} & \textbf{0.48}
& \textbf{19.66} & \textbf{1.42} \\

TC CO
& \textbf{2.15} & -0.25
& \textbf{9.22} & -0.53
& \textbf{17.06} & -1.57 \\

TC NO$_2$
& \textbf{3.40} & \textbf{0.16}
& \textbf{11.52} & \textbf{1.90}
& \textbf{19.92} & \textbf{3.56} \\

TC NO
& \textbf{0.30} & \textbf{0.35}
& \textbf{9.76} & \textbf{1.92}
& \textbf{17.69} & \textbf{4.46} \\

TC O$_3$
& \textbf{4.00} & -0.32
& \textbf{15.18} & -0.73
& \textbf{26.62} & -1.02 \\

TC SO$_2$
& \textbf{1.47} & \textbf{2.48}
& \textbf{7.82} & \textbf{11.45}
& \textbf{13.39} & \textbf{17.80} \\

CO (1000 hPa)
& \textbf{1.92} & \textbf{0.17}
& \textbf{6.06} & \textbf{0.26}
& \textbf{11.28} & \textbf{0.24} \\

NO (1000 hPa)
& -5.80 & \textbf{0.64}
& \textbf{1.13} & \textbf{1.26}
& \textbf{8.12} & \textbf{1.60} \\

NO$_2$ (1000 hPa)
& \textbf{2.63} & \textbf{0.22}
& \textbf{8.96} & \textbf{0.87}
& \textbf{16.31} & \textbf{1.33} \\

SO$_2$ (1000 hPa)
& -4.64 & \textbf{0.07}
& \textbf{4.61} & \textbf{0.73}
& \textbf{12.77} & \textbf{1.25} \\

O$_3$ (1000 hPa)
& \textbf{3.10} & \textbf{0.09}
& \textbf{12.66} & \textbf{0.62}
& \textbf{21.44} & \textbf{0.30} \\
\midrule
Mean
& \textbf{1.19} & \textbf{0.31}
& \textbf{8.42} & \textbf{1.45}
& \textbf{15.95} & \textbf{2.35} \\
\bottomrule
\end{tabular}
\end{table}

\clearpage
\subsection{Additional figures}
To ensure a fair comparison, AirMPA, Aurora, and CAMS are evaluated from the same pollutant initialization states. Specifically, the pollutant composition fields at $t_0-12\,\mathrm{h}$ and $t_0$ are taken from CAMS FC for all methods. After initialization, the models differ in their forecasting mechanisms and meteorological inputs: AirMPA uses FuXi-2.0 forecast fields as external forcing, Aurora follows its original coupled forecasting framework, and CAMS serves as the numerical baseline. This setting is consistent with operational air-quality forecasting, where composition initial fields are typically provided by an analysis or reanalysis product and then advanced forward by the forecasting system.

\begin{figure}[htbp]
\centering
\includegraphics[width=1\textwidth]{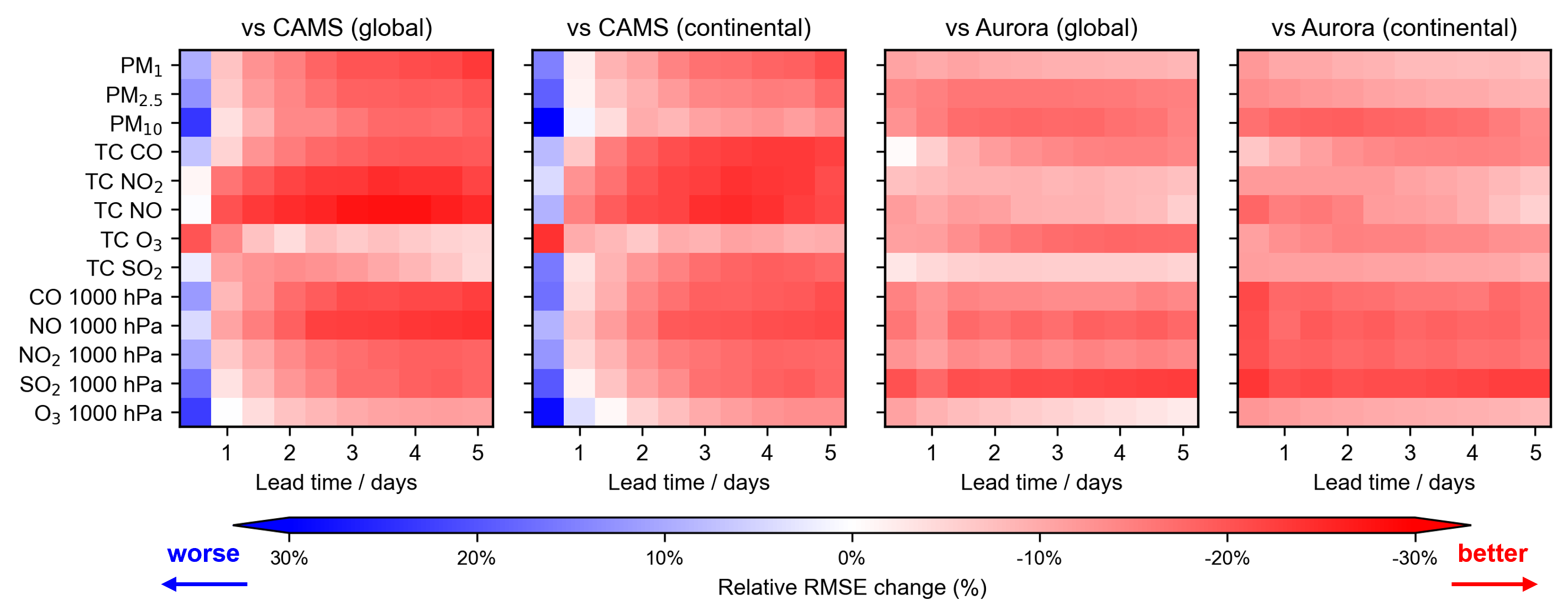}
\caption{Comparison of forecasting performance across AirMPA (with FuXi), CAMS, and Aurora.}
\label{fig:main_heatmap}
\end{figure}

Figure~\ref{fig:main_heatmap} presents a comprehensive performance evaluation across three modeling frameworks: AirMPA (driven by FuXi-2.0 forecasts), CAMS, and Aurora. Results are categorized by global and terrestrial domains, providing a rigorous assessment of model skill under both historical reanalysis and real-time operational forecasting scenarios.

\clearpage

\begin{figure}[t]
\centering
\includegraphics[width=1\textwidth]{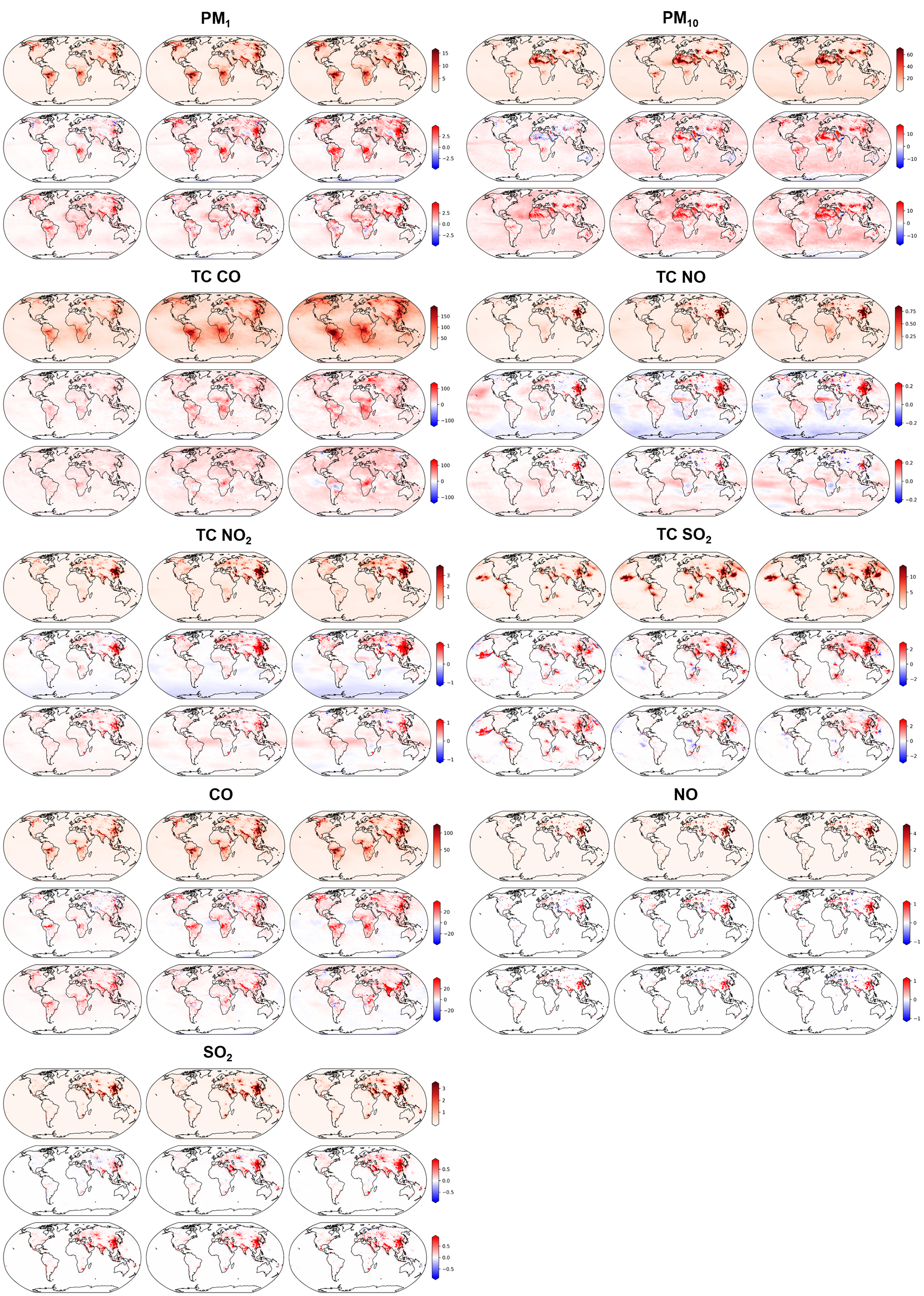}
\caption{Same as Figure~\ref{fig:countourf}, but for other pollutants. The three columns in each subplot correspond to lead times of 1, 3, and 5 days, while the rows correspond to AirMPA, CAMS – AirMPA, and Aurora – AirMPA, respectively.}
\label{fig:all_countourf}
\end{figure}

\clearpage

\begin{figure}[htbp]
\centering
\includegraphics[width=1\textwidth]{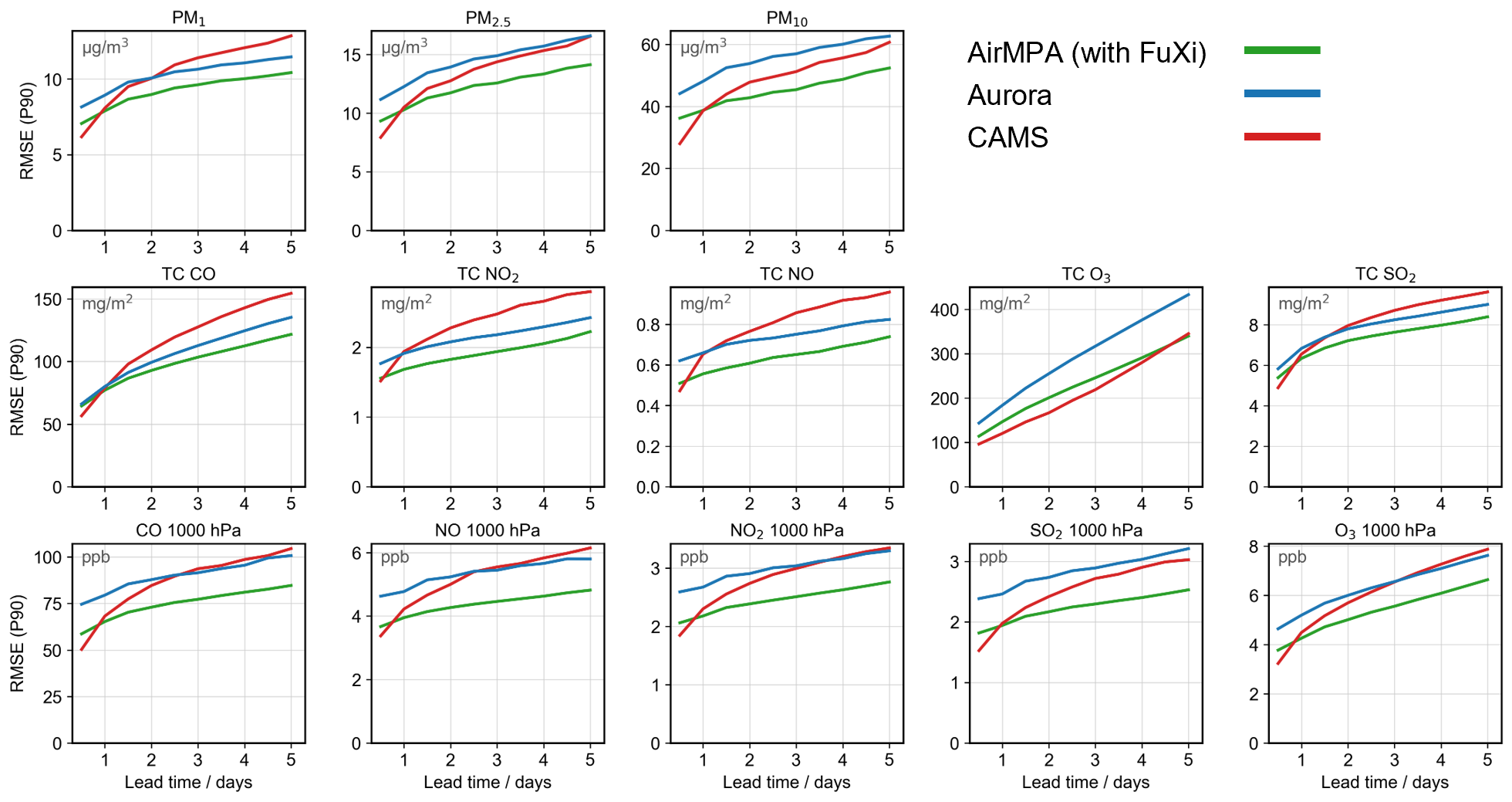}
\caption{Comparison of extreme forecasting performance(90\% RMSE during June 2022 -- November 2022) across AirMPA (with FuXi), CAMS, and Aurora.}
\label{fig:RMSE90}
\end{figure}

\begin{figure}[htbp]
\centering
\includegraphics[width=1\textwidth]{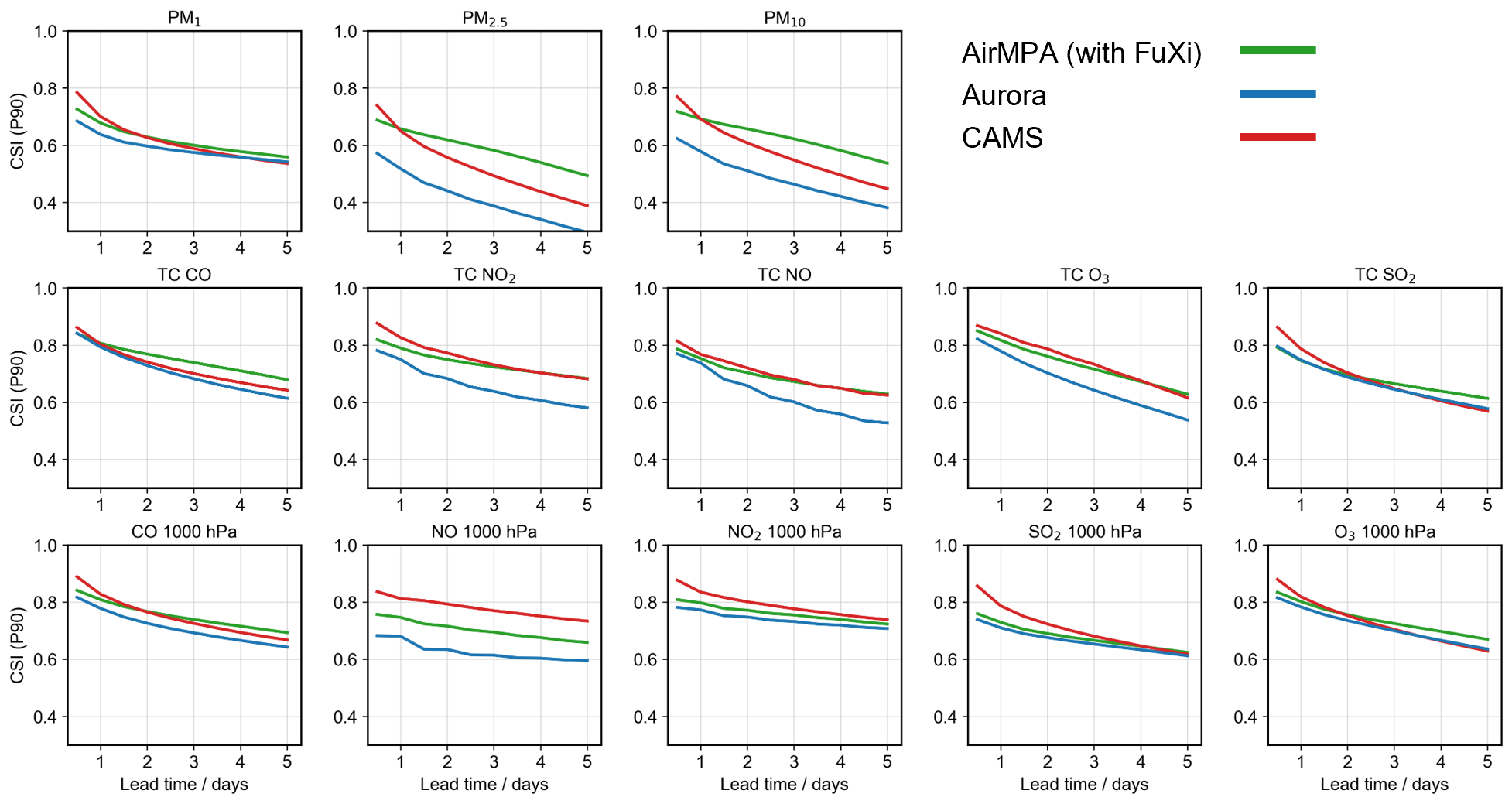}
\caption{Comparison of extreme forecasting performance(90\% CSI during June 2022 -- November 2022) across AirMPA (with FuXi), CAMS, and Aurora.}
\label{fig:CSI90}
\end{figure}

\begin{figure}[htbp]
\centering
\includegraphics[width=1\textwidth]{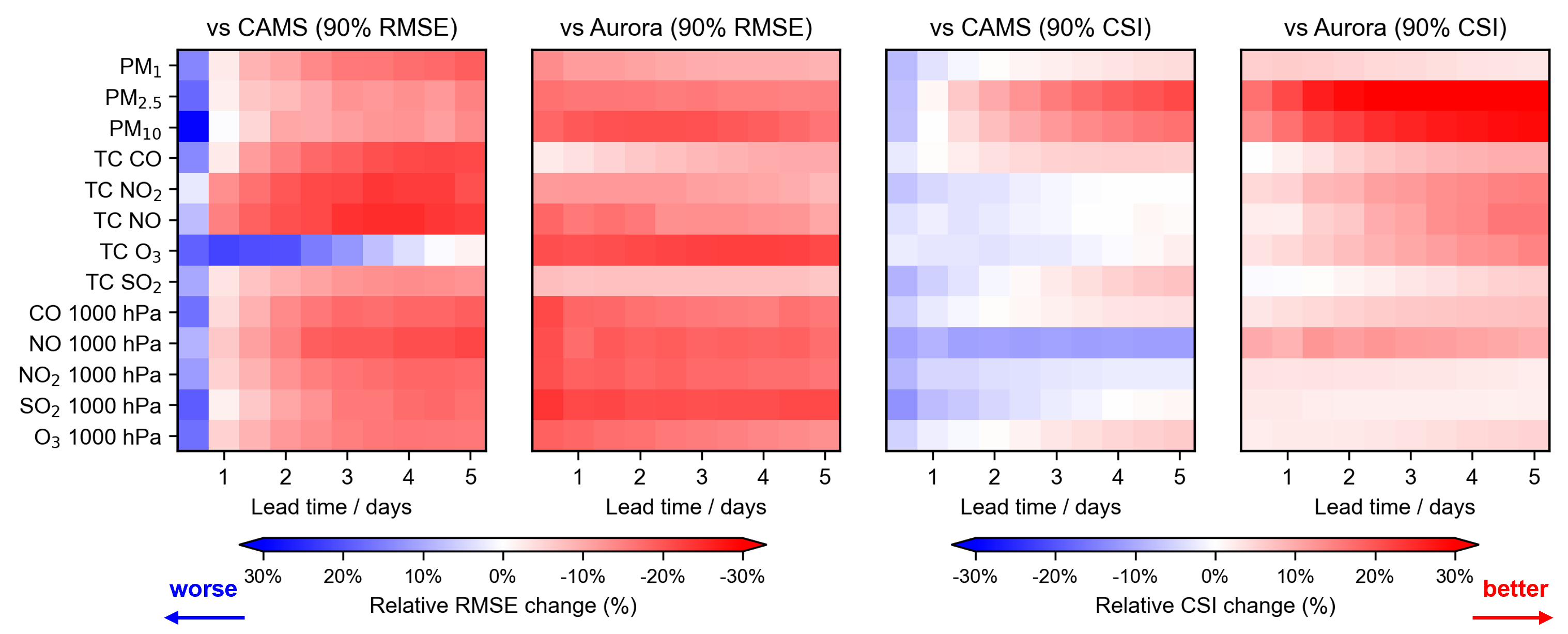}
\caption{Comparison of the AirMPA model in extreme value forecasting metrics relative to CAMS and Aurora.}
\label{fig:extreme_heatmap}
\end{figure}

Figures~\ref{fig:RMSE90}, \ref{fig:CSI90}, and \ref{fig:extreme_heatmap} illustrate the comparative predictive skill of AirMPA (driven by FuXi-2.0 forecasts), CAMS, and Aurora specifically for extreme pollution events.

\clearpage

\begin{figure}[htbp]
\centering
\includegraphics[width=1\textwidth]{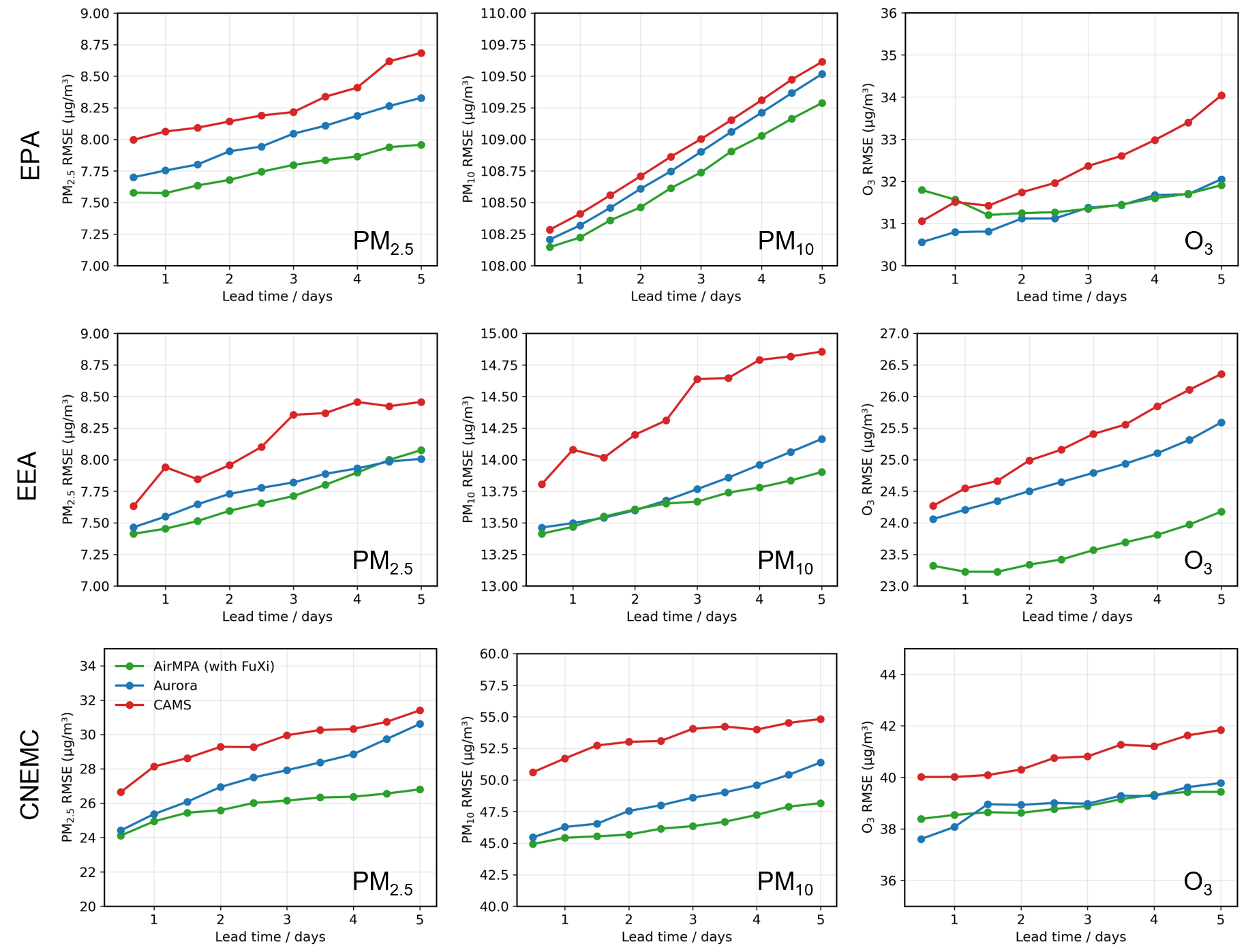}
\caption{Comparison of forecasting performance of AirMPA, Aurora, and CAMS at monitoring stations in the United States (EPA), Europe (EEA) and China (CNEMC).}
\label{fig:site_validation}
\end{figure}

\begin{figure}[htbp]
\centering
\includegraphics[width=1\textwidth]{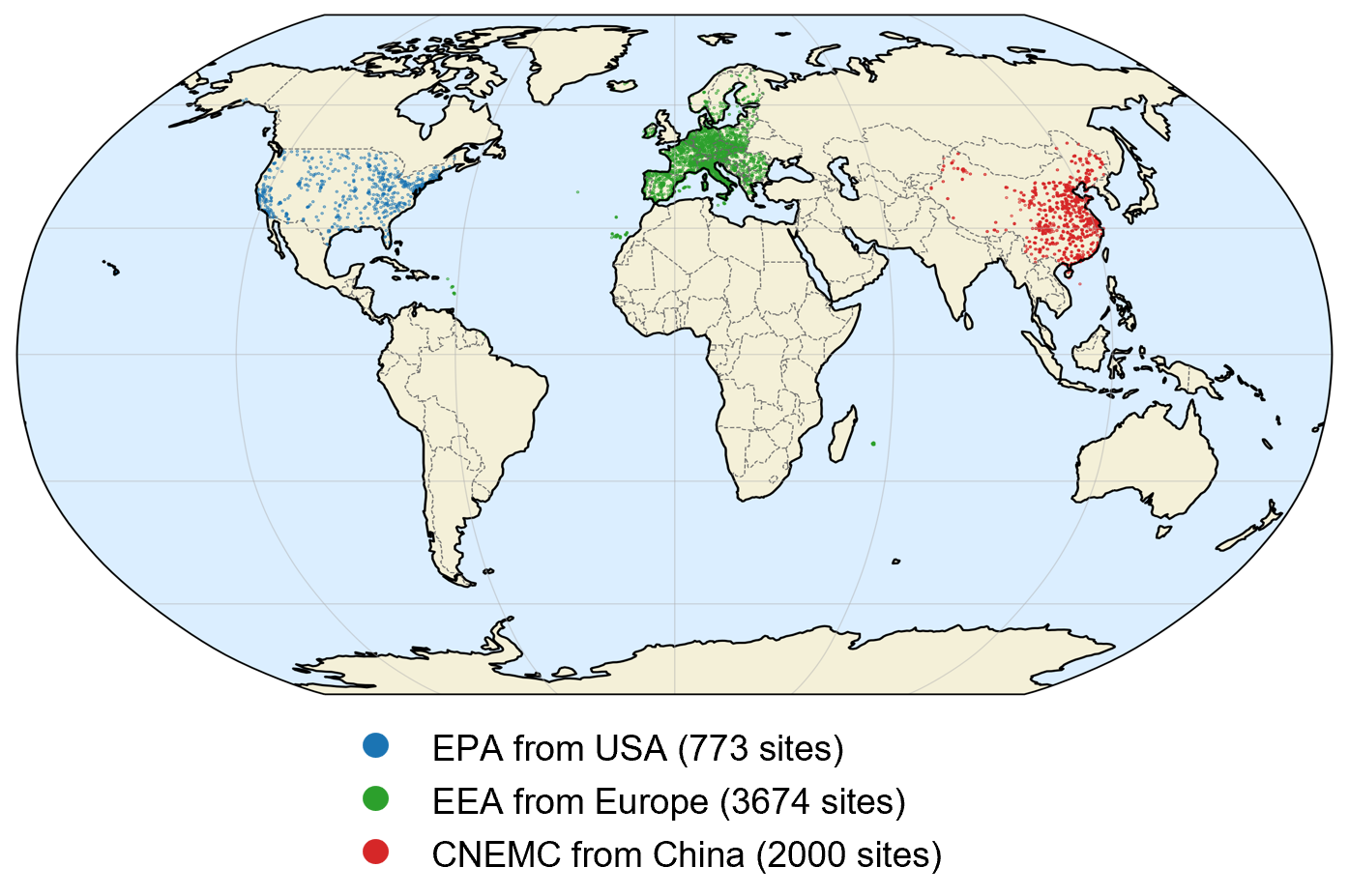}
\caption{Distribution of monitoring stations in the United States (EPA), Europe (EEA) and China (CNEMC).}
\label{fig:global_site}
\end{figure}

Figure~\ref{fig:site_validation} presents a performance comparison of three models—AirMPA, Aurora, and CAMS—at air quality monitoring stations operated by the U.S. Environmental Protection Agency (EPA)\footnote{\url{https://aqs.epa.gov/aqsweb/airdata/download_files.html}}, the European Environment Agency (EEA)\footnote{\url{https://www.eea.europa.eu/en/datahub/datahubitem-view/778ef9f5-6293-4846-badd-56a29c70880d}}, and the China National Environmental Monitoring Centre (CNEMC)\footnote{\url{https://www.cnemc.cn/}}. The spatial distribution of these stations is shown in Figure~\ref{fig:global_site}.

\clearpage

\section*{NeurIPS Paper Checklist}

\begin{enumerate}

\item {\bf Claims}
    \item[] Question: Do the main claims made in the abstract and introduction accurately reflect the paper's contributions and scope?
    \item[] Answer: \answerYes{} % Replace by \answerYes{}, \answerNo{}, or \answerNA{}.
    \item[] Justification: The abstract and introduction (section 1) have clearly outlined our primary contributions.

\item {\bf Limitations}
    \item[] Question: Does the paper discuss the limitations of the work performed by the authors?
    \item[] Answer: \answerYes{} % Replace by \answerYes{}, \answerNo{}, or \answerNA{}.
    \item[] Justification: We provide a comprehensive discussion of the model's limitations in Appendix C

\item {\bf Theory assumptions and proofs}
    \item[] Question: For each theoretical result, does the paper provide the full set of assumptions and a complete (and correct) proof?
    \item[] Answer: \answerNA{} % Replace by \answerYes{}, \answerNo{}, or \answerNA{}.
    \item[] Justification: this paper does not include theoretical results.

    \item {\bf Experimental result reproducibility}
    \item[] Question: Does the paper fully disclose all the information needed to reproduce the main experimental results of the paper to the extent that it affects the main claims and/or conclusions of the paper (regardless of whether the code and data are provided or not)?
    \item[] Answer: \answerYes{} % Replace by \answerYes{}, \answerNo{}, or \answerNA{}.
    \item[] Justification: We provide a comprehensive description of the model architecture and experimental pipeline in Sections 3 and 4, respectively. Furthermore, Appendix A explicitly details the data availability, including preprocessing steps and source links, ensuring that all information necessary to reproduce our main findings is fully disclosed.

\item {\bf Open access to data and code}
    \item[] Question: Does the paper provide open access to the data and code, with sufficient instructions to faithfully reproduce the main experimental results, as described in supplemental material?
    \item[] Answer: \answerYes{} % Replace by \answerYes{}, \answerNo{}, or \answerNA{}.
    \item[] Justification: The AirMPA code is publicly available at \url{https://github.com/yhWang0712/AirMPA}. Because the full source datasets are too large to redistribute with the repository, we provide a sample data subset for running a 3-day forecast. Appendix E lists the original data sources and access conditions. The sample supports an inference demonstration rather than reproduction of the full evaluation.

\item {\bf Experimental setting/details}
    \item[] Question: Does the paper specify all the training and test details (e.g., data splits, hyperparameters, how they were chosen, type of optimizer) necessary to understand the results?
    \item[] Answer: \answerYes{} % Replace by \answerYes{}, \answerNo{}, or \answerNA{}.
    \item[] Justification: As shown in Section 4.

\item {\bf Experiment statistical significance}
    \item[] Question: Does the paper report error bars suitably and correctly defined or other appropriate information about the statistical significance of the experiments?
    \item[] Answer: \answerNo{} % Replace by \answerYes{}, \answerNo{}, or \answerNA{}.
    \item[] Justification: We do not report multi-seed error bars because training global forecasting models at this scale is computationally expensive. Instead, we evaluate all methods on a continuous six-month independent test period across multiple lead times, pollutants, and spatial domains.

\item {\bf Experiments compute resources}
    \item[] Question: For each experiment, does the paper provide sufficient information on the computer resources (type of compute workers, memory, time of execution) needed to reproduce the experiments?
    \item[] Answer: \answerYes{} % Replace by \answerYes{}, \answerNo{}, or \answerNA{}.
    \item[] Justification: We specify this part in Section 4.2.3 and Appendix B.4.
    
\item {\bf Code of ethics}
    \item[] Question: Does the research conducted in the paper conform, in every respect, with the NeurIPS Code of Ethics \url{https://neurips.cc/public/EthicsGuidelines}?
    \item[] Answer: \answerYes{} % Replace by \answerYes{}, \answerNo{}, or \answerNA{}.
    \item[] Justification: The research conforms with the NeurIPS Code of Ethics.

\item {\bf Broader impacts}
    \item[] Question: Does the paper discuss both potential positive societal impacts and negative societal impacts of the work performed?
    \item[] Answer: \answerYes{} % Replace by \answerYes{}, \answerNo{}, or \answerNA{}.
    \item[] Justification: We discuss the broader impacts of the paper in Section D.
    
\item {\bf Safeguards}
    \item[] Question: Does the paper describe safeguards that have been put in place for responsible release of data or models that have a high risk for misuse (e.g., pre-trained language models, image generators, or scraped datasets)?
    \item[] Answer: \answerNA{} % Replace by \answerYes{}, \answerNo{}, or \answerNA{}.
    \item[] Justification: NA

\item {\bf Licenses for existing assets}
    \item[] Question: Are the creators or original owners of assets (e.g., code, data, models), used in the paper, properly credited and are the license and terms of use explicitly mentioned and properly respected?
    \item[] Answer: \answerYes{}
    \item[] Justification: We explicitly cite and credit all third-party datasets, models, and code used in this work. ERA5 and the CAMS products used in this study, including CAMS global reanalysis, CAMS global atmospheric composition forecasts, and CAMS global emission inventories, are distributed through the Copernicus Climate Data Store and Atmosphere Data Store under the CC-BY licence, and we comply with the required Copernicus attribution, license, and disclaimer terms. FuXi-2.0 forecast fields are used for academic research with authorization from the model developers; where public FuXi assets are involved, the corresponding code and model-weight terms are respected, including the Apache-2.0 licence for the public inference/plugin code and the CC BY-NC-SA 4.0 non-commercial terms for released FuXi model weights. Aurora is used only as a research benchmark; its source code and released checkpoints are credited to Microsoft and are distributed under the MIT license. All official repositories, data stores, and license or terms-of-use information are listed in the Data Availability section, and no third-party asset is redistributed beyond the scope permitted by its license or authorization.

\item {\bf New assets}
    \item[] Question: Are new assets introduced in the paper well documented and is the documentation provided alongside the assets?
    \item[] Answer: \answerYes{}
    \item[] Justification: We release the AirMPA implementation together with a sample data subset and instructions for running a 3-day forecast. The example illustrates the required input format and inference workflow without requiring users to download the full source datasets. The model's scope and limitations are discussed in Appendix C.

\item {\bf Crowdsourcing and research with human subjects}
    \item[] Question: For crowdsourcing experiments and research with human subjects, does the paper include the full text of instructions given to participants and screenshots, if applicable, as well as details about compensation (if any)? 
    \item[] Answer: \answerNA{} % Replace by \answerYes{}, \answerNo{}, or \answerNA{}.
    \item[] Justification: This work does not involve crowdsourcing experiments or research with human subjects.

\item {\bf Institutional review board (IRB) approvals or equivalent for research with human subjects}
    \item[] Question: Does the paper describe potential risks incurred by study participants, whether such risks were disclosed to the subjects, and whether Institutional Review Board (IRB) approvals (or an equivalent approval/review based on the requirements of your country or institution) were obtained?
    \item[] Answer: \answerNA{} % Replace by \answerYes{}, \answerNo{}, or \answerNA{}.
    \item[] Justification: This work does not involve human-subject experiments or personally identifiable data, so IRB approval is not applicable.

\item {\bf Declaration of LLM usage}
    \item[] Question: Does the paper describe the usage of LLMs if it is an important, original, or non-standard component of the core methods in this research? Note that if the LLM is used only for writing, editing, or formatting purposes and does \emph{not} impact the core methodology, scientific rigor, or originality of the research, declaration is not required.
    %this research? 
    \item[] Answer: \answerNA{} % Replace by \answerYes{}, \answerNo{}, or \answerNA{}.
    \item[] Justification: LLMs are not used as an important, original, or non-standard component of the core methodology.

\end{enumerate}

\end{document}